\documentclass[amsmath,amssymb,aps,pre,showkeys,reprint,eqsecnum]{revtex4-1}

\usepackage{graphicx}  % Include figure files
\usepackage{dcolumn}   % Align table columns on decimal point
\usepackage{bm}        % Bold math

\begin{document}

\title{Passive advection of a vector field: Anisotropy, finite correlation
time, exact solution and logarithmic corrections to ordinary scaling}

\author{N. V. Antonov and N. M. Gulitskiy}

\email{n.antonov@spbu.ru, ngulitskiy@gmail.com}

\affiliation{Chair of High Energy Physics and Elementary Particles \\
Department of Theoretical Physics, Faculty of Physics \\
Saint Petersburg State University, Ulyanovskaja~1 \\
Saint~Petersburg--Petrodvorez, 198504 Russia}

\begin{abstract}
In this work we study the generalization of the problem, considered in
[{\it Phys. Rev. E} {\bf 91}, 013002 (2015)], to the case of {\it finite}
correlation time of the environment (velocity) field.
The model describes a vector (e.g., magnetic) field,
passively advected by a strongly anisotropic turbulent flow.
Inertial-range asymptotic behavior is studied by means of the field
theoretic renormalization group and the operator product expansion.
The advecting velocity field is Gaussian, with finite correlation time and
preassigned pair correlation function.
Due to the presence of distinguished direction ${\bf n}$, all the multiloop
diagrams in this model are vanish, so that the results obtained are exact.
The inertial-range behavior of the model is described by two regimes
(the limits of vanishing or infinite correlation time) that correspond
to the two nontrivial fixed points of the RG equations. Their stability
depends on the relation between the exponents in the energy spectrum
${\cal E} \propto  k_{\bot}^{1-\xi}$ and the dispersion law
$\omega \propto  k_{\bot}^{2-\eta}$.
In contrast to the well known isotropic Kraichnan's model, where various
correlation functions exhibit anomalous scaling behavior with infinite sets
of anomalous exponents, here the corrections to ordinary scaling are
polynomials of logarithms of the integral turbulence scale $L$.
\end{abstract}



\maketitle

%%%%%%%%%%%%%%%%%%%%%%%%%%%%%%%%%%%%%%%%%%%%%%%%%%%%%%%%%%%%%%%%%%%%%%%%%%%%%%%%%%%%%%%%%%%%%%%%%%%%%%%%%%%%%%%%%%%%%%%%%%%%%%%%%%%%%%%%%%%%%%%%%%%%%%%%%%%%%%%%%%%%%%%%
\section{Introduction} \label{sec:Intro}

Over decades much attention has been paid to the problem of intermittency
and anomalous scaling in fully developed turbulence.
Both the natural experiments and numerical simulations suggest that the
violation of the classical Kolmogorov--Obukhov theory~\cite{Legacy}
is even more strongly pronounced for a advected field than for the velocity
field itself; see, e.g.,~\cite{Advection,FGV}
and references therein. At the same time, the problem of passive advection
appears to be easier tractable theoretically.
Although the theoretical description of the fluid turbulence on the basis of
the stochastic Navier--Stokes (NS) equations remains essentially an open
problem, considerable progress has been achieved in understanding
passive advection by random ``synthetic'' velocity fields.
The most remarkable progress has been achieved for the so-called
Kraichnan's rapid-change model~\cite{Kraich1}, in which the velocity
field is modeled by Gaussian ensemble, not correlated in time,
with zero mean and pair correlation function of the form
\begin{widetext}
\begin{equation}
\label{Kraich}
\langle v_{i}(x) v_{j}(x')\rangle = \delta(t-t')\, D_{0}\
\int_{k>m} \frac{d{\bf k}}{(2\pi)^{d}} \, P_{ij}({\bf k}) \,
 \frac{1}{k^{d+\xi}}\ e^{{\rm i}{\bf k}\cdot({\bf x}-{\bf x'})}.
\end{equation}
\end{widetext}
Here $P_{ij}({\bf k}) = \delta _{ij} - k_i k_j / k^2$ is the transverse
projector, $k\equiv |{\bf k}|$, $D_{0}>0$ is an amplitude factor, $d$ is the
dimensionality of the ${\bf x}$ space and $0<\xi<2$ is a parameter with the
real (``Kolmogorov'') value $\xi=4/3$.
For the first time, the anomalous exponents have been calculated on the
basis of a microscopic model and within
regular expansions in formal small parameters~\cite{FGV}.

A passively advected field may be chosen both scalar and vector, the latter
case corresponds to the magnetohydrodynamic (MHD) turbulence. From the
experimental point of view it is a special problem, closely related to
the processes taken place in solar corona, e.g., with solar wind;
for detailed discussion see~\cite{GM-mod,E,SW2} and references therein.

In solar flares, highly energetic and anisotropic large-scale motions coexist
with small-scale coherent structures, finally responsible for the dissipation.
A simplified description of the situation was proposed in~\cite{E}: the
large-scale field $B^{0}_{i} = n_{i} B^{0}$ dominates the dynamics in the
distinguished direction ${\bf n}$, while the activity in the perpendicular
plane is described as nearly two-dimensional.

The observations and simulations show that the scaling behavior in the solar
wind is closer to the anomalous scaling
of the three-dimensional fully developed hydrodynamic turbulence,
rather than to simple Iroshnikov-Kraichnan scaling suggested by the
two-dimensional picture with the inverse energy cascade~\cite{SW2}.
Thus, further analysis of more realistic three-dimensional models is welcome.

One of the possibilities to make original Kraichnan's model~\eqref{Kraich}
anisotropic is to replace the ordinary transverse projector with
the tensor quantity $T_{ij}({\bf k})$, which contains a fixed unit
vector ${\bf n}$:
\begin{equation}
\label{Anis-Old}
T_{ij}({\bf k}) = a(\psi)P_{ij}({\bf k})
+ b(\psi)n_sn_lP_{is}({\bf k})P_{jl}({\bf k}),
\end{equation}
where $a(\psi)$ and $b(\psi)$ are some functions of $\psi$,
the angle between the vectors ${\bf n}$ and ${\bf k}$; see,
e.g.,~\cite{Uni,Uni2,Uni3}. This
formulation of the problem corresponds to the small-scale anisotropy and
contains an isotropic model as a special case,
if $a(\psi)=1$ and $b(\psi)=0$.

Another possibility is the ``strongly anisotropic'' model that does not
contain an isotropic one as
a special case and is obtained by introducing the velocity field ${{\boldsymbol{v}}}$
having preferred direction ${\bf n}$:
\begin{equation}
\label{V-n}
\boldsymbol{v}(t,{\bf x})={\bf n}\times v(t,\ {\bf x_\perp}).
\end{equation}

In this paper, we consider a more realistic model with {\it finite}
(and not small) correlation time. For this purpose the correlation
function~\eqref{Kraich} has to be modified, and instead of a constant,
which is Fourier transform of $\delta(t-t')$, in the frequency space it
becomes a function of $\omega$.
In common cases this modification disrupts the Galilean
invariance~\cite{FinTime-GalileanLack} and is interesting only as a model,
but in the presence of the anisotropy Galilean invariance survives and the
model is invariant under some special Galilean transformations
(more precisely see below).

The energy spectrum of the velocity in the inertial range has the form
${\cal E} \propto  k_{\bot}^{1-\xi}$, while the correlation time
at the momentum $k$ scales as $k^{-2+\eta}$.
Such ensemble was employed in some models, studied
in~\cite{FinTime,FinTimeEta}. It was shown that, depending on the values of
the
exponents $\xi$ and $\eta$, the model reveals various types of
inertial-range scaling regimes with nontrivial anomalous exponents, which
were explicitly derived to the first~\cite{FinTime} and
second~\cite{FinTimeEta}
orders of the double expansion in $\xi$ and $\eta$.

It is necessity to stress, that
the Kraichnan's model~\eqref{Kraich} and its generalizations
correspond to passive field approximation: if we neglect the
influence of advected field $\boldsymbol{\theta}$ to the dynamics
of the environment (velocity) field ${{\boldsymbol{v}}}$,
the latter can be modeled by statistical ensembles
with prescribed properties.
This approximation is valid when the
gradients of the magnetic fields are not too large.

A most powerful method to study the anomalous scaling in various statistical
models
of turbulent advection provided by the field theoretic renormalization
group (RG) and operator product expansion (OPE); see
the monographs~\cite{Zinn,Vasiliev-Green} and references therein.
In the RG+OPE
scenario~\cite{RG}, anomalous scaling emerges as a consequence of the
existence in the model of composite fields (``composite operators'' in
the quantum-field terminology) with {\it negative} scaling dimensions;
see~\cite{JphysA} for a review and the references.
In a number of papers the RG+OPE approach
was applied to the case of passive vector (magnetic) fields in Kraichnan's
ensemble, and to its generalizations (large-scale anisotropy, helicity,
compressibility, finite correlation time, non-Gaussianity, more general form
of the nonlinearity);
see~\cite{Lanotte2-mod,AntGul2012-mod,Marian,Marian2,Kotumay}
and references therein.
Explicit analytical expressions were derived for the
anomalous exponents to the first~\cite{Lanotte2-mod} and the second
\cite{AntGul2012-mod,Marian} orders in $\xi$.
For the pair correlation function of the magnetic field, exact results
were obtained within the zero-mode approach~\cite{V96-mod}.

In this paper, we apply the field theoretic renormalization group and
operator product expansion to the inertial-range behavior
of strongly anisotropic
MHD turbulence within the framework of a simplified model,
which corresponds to the problem of a passive vector field advected
by the Gaussian
ensemble with prescribed statistics.
The velocity field ${{\boldsymbol{v}}}$ is chosen to be oriented
along a fixed direction ${\bf n}$ (``orientation of a large-scale flare''
in the context of the solar corona dynamics) and depends only on the
coordinates in the subspace orthogonal to ${\bf n}$. In the momentum space,
its correlation function is
some function of $k_{\bot}$ and frequency $\omega$, where
$k_{\bot}= |{\bf k}_{\bot}|$ and ${\bf k}_{\bot}$ is the component of the
momentum (wave number) ${\bf k}$ perpendicular to ${\bf n}$.
This model can be viewed as
a $d$-dimensional generalization of the strongly anisotropic velocity
ensemble introduced in~\cite{AM} in connection with the turbulent diffusion
problem and further studied and generalized in a number of
papers~\cite{AM1-mod,AM5-mod,Glimm-mod,Foba,AntMal2011}.

The advecting equation for the
passive field $\boldsymbol{\theta}$ involves a general relative coefficient
${\cal A}$, which unifies different physical situations: the kinematic
MHD model, the linearized NS equation and the passive
admixture with complex internal structure of the particles.

In~\cite{AntMal2011} the problem of anomalous scaling in the higher-order
correlation functions of
a {\it scalar} field, advected by such a velocity ensemble, was studied by
the RG+OPE techniques. It was shown that there exists some set of fixed
points, which governs infrared (IR) behavior of the system.
Another conclusion of that work is that in sharp
contrast to the isotropic Kraichnan's model and its numerous descendants,
due to the mixing of
{\it families} of relevant composite operators the
correlation functions show no anomalous scaling and have finite limits when
the integral turbulence scale tends to infinity.

Further modification of that problem, namely
advection of the {\it vector} field by {\it decorrelated} in time velocity
field, was studied in~\cite{VectorN}. In contrast to~\cite{AntMal2011}, the
inertial-range behavior
of vector fields appears to be even more exotic: instead
of power-like anomalies, there are logarithmic corrections to ordinary
scaling, determined by naive (canonical) dimensions.

The main result of the present paper is that the inertial-range behavior
of vector fields advected by velocity ensemble with finite correlation
time combines both the above features:
as in the scalar case, there is a set of fixed points, governing the
IR behavior;
as in the zero-time correlation model, the inertial-range behavior of
vector fields has logarithmic corrections to ordinary scaling.
The key point is that
the matrices of scaling dimensions (``critical dimensions'' in the
terminology of the theory of critical state) of the relevant families
of composite operators appear nilpotent and cannot be diagonalized.
They can only be brought to Jordan form; hence the logarithms.

Another interesting property, inherited from the zero-time correlation model,
is that all multiloop diagrams are equal to zero and therefore the set of
fixed points and the existence of logarithmic corrections are proven exactly.
Moreover, in contrast to previous one, this model has {\it two} types of such
nontrivial diagrams, with different causes to be equal to zero.
The physical meaning of this feature is not yet clarified, but it is
clear that it is closely connected with the presence of the anisotropy vector
${\bf n}$.

The paper is organized as follows.

In Sec.~\ref{sec:Model} we give a detailed description of the model.
In Sec.~\ref{sec:QFT} we present the field theoretic formulation
of the model and the corresponding diagrammatic techniques.
In Sec.~\ref{sec:Renorm} we establish renormalizability of the model
and derive explicit exact expressions for the renormalization constants
and RG functions (anomalous dimensions and $\beta$-functions).
Due to the presence of the anisotropy,
the linear response function, the only Green
function in the model that contains superficial ultraviolet (UV)
divergences,
is given exactly by the one-loop approximation.

It is shown that the IR behavior of the model is confined with only
two limiting cases: the rapid-change type behavior and the ``frozen''
(time-independent) behavior.
In contrast to the isotropic case, where
the physical (Kolmogorov) point $\xi=8/3$, $\eta=4/3$ lies exactly
on the crossover line between the rapid-change and frozen
regimes~\cite{FinTime,FinTimeEta,Chetak}, now this point lies deep inside
the domain of
stability of the nontrivial rapid-change behavior; there is no crossover
line going through this point. This result is in agreement with
the exact analysis of the $d=(1+1)$-dimensional case~\cite{Glimm-mod}
and in disagreement with~\cite{AM,AM1-mod}.

The corresponding differential equations of IR scaling are derived,
with the exactly known critical dimensions.

In Sec.~\ref{sec:Ops1} we discuss the renormalization of composite
operators and present explicit expressions for the matrices of
anomalous dimensions and critical dimensions.
It is shown that these matrices are given exactly by the one-loop
approximation. The matrices of anomalous dimensions appear to be nilpotent.
As a result, the IR behavior of the pair correlation functions
of the composite operators is given by canonical powers, corrected by
polynomials of logarithms. To obtain inertial-range behavior
we have to combine this result with the corresponding OPE's.
Finally, asymptotic behavior of the pair correlation functions involves two
types of large logarithms, where the separation enters with the typical
UV and IR scales (dissipation scale and integral scale).

Sec.~\ref{sec:Conc} is reserved for conclusions.

%%%%%%%%%%%%%%%%%%%%%%%%%%%%%%%%%%%%%%%%%%%%%%%%%%%%%%%%%%%%%%%%%%%%%%%%%%%%%%%%%%%%%%%%%%%%%%%%%%%%%%%%%%%%%%%%%%%%%%%%%%%%%%%%%%%%%%%%%%%%%%%%%%%%%%%%%%%%%%%%%%%%%%%%
\section{Description of the model} \label{sec:Model}
%%%%%%%%%%%%%%%%%%%%%%%%%%%%%%%%%%%%%%%%%%%%%%%%%%%%%%%%%%%%%%%%%%%%%%%%%%%%%%%%%%%%%%%%%%%%%%%%%%%%%%%%%%%%%%%%%%%%%%%%%%%%%%%%%%%%%%%%%%%%%%%%%%%%%%%%%%%%%%%%%%%%%%%%

If the field $\boldsymbol{v}$ is chosen in the strongly anisotropic form
(\ref{V-n}), the turbulent advection of a passive vector field
$\boldsymbol{\theta}(x)\equiv
\boldsymbol{\theta}(t,{\bf x})$ is described by the
stochastic equation~\cite{VectorN,Ant-Hnat-Hon-Jut-11}
\begin{equation}
\label{Adv}
\partial_t \theta_i+\partial_k\left(v_k\theta_i-{\cal A}_0\ v_i
\theta_k\right)+\partial {\cal P}=\nu_0\
(\partial_\perp^2+f_0\partial_\parallel^2)\theta_i+f_i,
\end{equation}
where $\theta_i(x)$ is a vector field,
$x\equiv\left\{t,{\bf x}\right\}$,
$\partial _t \equiv \partial /\partial t$,
$\partial _i \equiv \partial /\partial x_{i}$,
${\bf n}$ is a unit vector that determines the distinguished direction,
${\bf x}_{\bot}$ and $\boldsymbol{\partial}_\perp$ are the components of the
vectors ${\bf x}$ and $\boldsymbol{\partial}$ perpendicular to ${\bf n}$,
$\partial_\parallel\equiv\boldsymbol{\partial}\cdot{\bf n}$,
$\nu _0$
is the molecular diffusivity coefficient, $\partial^{2}$ is the Laplace
operator, $\boldsymbol{v}(x)\equiv \{v_{i}(x)\}$
is the velocity field, $f_i\equiv f_i(x)$ is an
artificial Gaussian scalar noise with zero mean and correlation function
\begin{equation}
\label{Cik}
\left\langle f_i(t,\ {\bf x})\ f_k(t',\ {\bf x'})\right\rangle
=\delta(t-t')\ C_{ik}({\bf r}/L).
\end{equation}
Here ${\bf r=x-x'}$,
$r=\left|{\bf r}\right|$, the parameter $L\equiv M^{-1}$ is the
integral (external) turbulence scale related to the stirring, and $C_{ik}$
is a dimensionless function finite for $r/L\to0$ and
rapidly decaying for $r/L\to\infty$.

Both  $\boldsymbol{v}$ and ${\boldsymbol \theta}$ are divergence-free
(``solenoidal'') vector fields:
\begin{equation}
\label{Trans-Def}
\partial_{i}v_{i}=0, \quad \partial_{i}\theta_{i}=0.
\end{equation}

Following~\cite{amodel}, we included into the stochastic advection-diffusion
equation~(\ref{Adv}) additional arbitrary dimensionless parameter
${\cal A}_0$, which unifies different physical situations:
the case ${\cal A}_0=1$ corresponds to the kinematic MHD equation,
describing, for example, the evolution of the fluctuating part
${\boldsymbol \theta}\equiv {\boldsymbol \theta}(x)$ of the magnetic field
in the presence of a mean component ${\boldsymbol \theta}^0$, which is
supposed to be varying on a very large scale; the case ${\cal A}_0=-1$
corresponds to the linearization of the NS equation
around the rapid-change background velocity field; in the case
${\cal A}_0=0$ equation~\eqref{Adv} loses the stretching term
$\partial_k(v_i\theta_k)$ and the model
acquires additional symmetry under translations
${\boldsymbol \theta}\to{\boldsymbol \theta+const}$.
This case has to be studied separately, see~\cite{Matraz}.

The pressure term can be expressed as the solution of the Poisson equation
\begin{equation}
\partial^{2} {\cal P} = ({\cal A}_{0} - 1) \,
\partial_{i} v_{k}  \partial_{k} \theta_{i}
\label{Pois}
\end{equation}
and is needed to reconcile dynamics of the field $\theta_i$ with
transversality condition~\eqref{Trans-Def}.

For renormalizability reasons it is necessary to introduce additional
dimensionless constant $f_0$, which breaks the $O_d$ symmetry
of the Laplace operator to
$O_{d-1}\otimes Z_2$: $\partial^2\rightarrow\
\partial_\perp^2+f_0\partial_\parallel^2$
($Z_{2}$ is the reflection symmetry $x_{\parallel} \to - x_{\parallel}$).
Interpretation of the splitting of the Laplacian term can
be twofold; cf.~\cite{AntMal2011}.
On one hand, stochastic models of the type~(\ref{Adv}) must
include all
the IR relevant terms allowed by the symmetry, therefore
it is natural to
include the general value $f_{0}\ne1$ to the model from the very beginning.
On the other hand, the extension of the model to the case $f_{0}\ne1$ can be
viewed as a purely technical trick which is only needed to ensure the
multiplicative renormalizability and to derive the RG equations.

Instead of the real problem,  where the velocity field $\boldsymbol{v}(x)$
has to satisfy the NS equation with some additional terms that describe
the feedback of the advected field $\boldsymbol{\theta}(x)$ on the velocity
field, we will consider the {\it kinematic} problem, where
the reaction of the field $\boldsymbol{\theta}(x)$ on the velocity field
$\boldsymbol{v}(x)$ is neglected. It is assumed that, if the gradients of
$\boldsymbol{\theta}(x)$ are not too large, it does not affect essentially
dynamics of the conducting fluid. Thus, the field $\boldsymbol{v}(x)$ can
be simulated by statistical ensemble with prescribed statistics.
It is assumed to be Gaussian, strongly anisotropic [see~(\ref{V-n})],
homogeneous, with zero mean and a correlation
function~\cite{FinTime,FinTimeEta,AntMal2011}
\begin{equation}
\label{ViVk}
\left\langle v_i(t,\ {\bf x})\ v_k(t',\ {\bf x'})\right\rangle
=n_in_k\ \left\langle v(t,\ {\bf x_\perp})\ v(t',\ {\bf x'_\perp})
\right\rangle,
\end{equation}
where
\begin{equation}
\label{VV}
\left\langle v(t,\ {\bf x_\perp})\ v(t',\ {\bf x'_\perp})\right\rangle
= \int_{k>m}\frac{d{\bf k}}{(2\pi)^d}\ e^{i{\bf k\cdot(x-x')}}\
D_v(\omega,k).
\end{equation}
The function $D_v$ is chosen in the form
\begin{equation}
\label{Dv}
D_v(\omega,k)=2\pi\delta(k_\parallel)\ D_0\
\frac{k_\perp^{5-d-(\xi+\eta)}}{\omega^2+\left[\alpha_0\nu_0k_
\perp^{2-\eta}\right]^2}.
\end{equation}
Here $d$ is the dimensionality of the ${\bf x}$ space,
$k_\perp \equiv |{\bf k_\perp}|$, $1/m$ is another integral turbulence scale,
related to the stirring,
$D_0>0$ is an amplitude factor and symbol $k_\parallel$ denotes
the scalar product ${\bf k\cdot n}$. The function~\eqref{VV} involves
two independent exponents $\xi$ and $\eta$, which in the RG
approach play the role of two formal expansion parameters;
a new parameter $\alpha_0$ is needed for the dimensionality reason.
Depending of this parameter, the function~\eqref{Dv} demonstrates two
interesting
limiting cases: if $\alpha_0\to0$, $D_v(\omega)\propto\delta(\omega)$,
so that from the physs point of view this situation corresponds to the
independent of time (``frozen'') velocity field.
The situation $\alpha_0\to\infty$ in fact means that
$(\alpha_0\nu_0)^2\gg\omega^2$, so that this case corresponds to the
rapid-change model.

The relations
\begin{equation}
\label{D0}
D_0/\nu_0^3 f_0=\tilde{g}_0\equiv\Lambda^{\xi+\eta}
\end{equation}
define the coupling constant $\tilde{g}_0$, which plays the role of the
expansion parameter in the ordinary perturbation theory, and the
characteristic UV momentum scale $\Lambda$.

%%%%%%%%%%%%%%%%%%%%%%%%%%%%%%%%%%%%%%%%%%%%%%%%%%%%%%%%%%%%%%%%%%%%%%%%%%%%%%%%%%%%%%%%%%%%%%%%%%%%%%%%%%%%%%%%%%%%%%%%%%%%%%%%%%%%%%%%%%%%%%%%%%%%%%%%%%%%%%%%%%%%%%%%
\section{Field theoretic formulation of the model}\label{sec:QFT}
%%%%%%%%%%%%%%%%%%%%%%%%%%%%%%%%%%%%%%%%%%%%%%%%%%%%%%%%%%%%%%%%%%%%%%%%%%%%%%%%%%%%%%%%%%%%%%%%%%%%%%%%%%%%%%%%%%%%%%%%%%%%%%%%%%%%%%%%%%%%%%%%%%%%%%%%%%%%%%%%%%%%%%%%

%%%%%%%%%%%%%%%%%%%%%%%%%%%%%%%%%%%%%%%%%%%%%%%%%%%%%%%%%%%%%%%%%%%%%%%%%%%%%%%%%%%%%%%%%%%%%%%%%%%%%%%%%%%%%%%%%%%%%%%%%%%%%%%%%%%%%%%%%%%%%%%%%%%%%%%%%%%%%%%%%%%%%%%%
\subsection{The action functional and the Galilean symmetry} \label{sec:Diag}

The stochastic problem~(\ref{Adv})~--~(\ref{Dv}) is equivalent to the
field theoretic model of the extended set of three fields
$\Phi\equiv\left\{\boldsymbol{\theta},\boldsymbol{\theta}',
\boldsymbol{v}\right\}$ with the action functional
\begin{widetext}
\begin{equation}
\label{Action}
{\cal S}(\Phi)= -\frac{1}{2}v_iD_v^{-1}v_k+
\frac{1}{2}\theta'_i D_{\theta}\theta'_k
+\theta'_k\left[-\partial_t\theta_k-(v_i\partial_i)\theta_k
+{\cal A}_0(\theta_i\partial_i)v_k+\nu_0(\partial_\perp^2
+f_0\partial_\parallel^2)\theta_k\right] .
\end{equation}
\end{widetext}
Here all the terms, with the exception of the first one, represent the
De Dominicis--Janssen action for the stochastic
problem~(\ref{Adv}),~(\ref{Cik}) at fixed $\boldsymbol{v}$,
while the first term
represents the Gaussian averaging over $\boldsymbol{v}$. Furthermore,
$D_{\theta}$
and $D_{v}$ are the correlators~(\ref{Cik}) and~(\ref{ViVk}) respectively;
the needed integrations over $x=(t,{\bf x})$ and summations over
the vector indices are implied.

As a rule, synthetic velocity ensembles with finite correlation time
suffer from the lack of Galilean invariance, which can lead to some physical
pathologies; see, e.g., the discussion in~\cite{FinTime-GalileanLack}.
Surprisingly enough, the presence of the anisotropy can improve the
situation.

Indeed, it is directly checked that
in our strongly anisotropic case the action functional
(\ref{Action}) with the correlator (\ref{ViVk}) in its first term appears
invariant with respect to the Galilean transformation of a special form:
\begin{eqnarray}
\theta(t,{\bf x}) &\to& \theta(t,{\bf x}+{{\boldsymbol{u}}}t), \quad
\theta'(t,{\bf x}) \to \theta'(t,{\bf x}+{{\boldsymbol{u}}}t), \nonumber \\
{{\boldsymbol{v}}}(t, {\bf x}) &\to& {{\boldsymbol{v}}}
(t, {\bf x} +{{\boldsymbol{u}}}t) - {{\boldsymbol{u}}}.
\label{GT}
\end{eqnarray}
Here the transformation parameter has the form ${{\boldsymbol{u}}}={\bf n} u$ with the
vector ${\bf n}$ from (\ref{V-n}), so that the scalar coefficient in
(\ref{V-n}) changes as $v(t, {\bf x}_{\bot}) \to v(t, {\bf x}_{\bot})-u$
and the arguments ${\bf x}_{\bot}$ of all the fields in (\ref{GT}) remain
intact.

This fact can be interpreted as follows. Consider the generalized stochastic
NS equation
\begin{equation}
\partial_{t} v_{i} + (v_{l} \partial_{l}) v_{i} + \partial_{i} \wp
= R v_{i} + \phi_{i},
\label{NS}
\end{equation}
where $R$ is some differential operation acting only on spatial coordinates
and $\wp= - \partial^{-2} (\partial_{i}v_{l})(\partial_{l} v_{i})$ is the
pressure.  If the random force $\phi_{i}$ is taken to be white in time,
the equation (\ref{NS}) is Galilean covariant because it involves the full
covariant derivative $\partial_{t}+(v_{l} \partial_{l})$.

However, for the velocity field of the form (\ref{V-n}) all the nonlinear
terms in (\ref{NS}) vanish due to the independence
of the scalar coefficient $v$ on $x_{\parallel}$:
$ v_{k}\partial_{k} v_{i} = n_{i} v \partial_{\parallel} v =0$,
and similarly for the pressure. Thus the equation (\ref{NS}) becomes
in fact linear and generates a Gaussian velocity field.
Its pair correlation function has the form
\begin{equation}
\left\langle v_iv_j \right\rangle =
\frac{D^\phi_{ij}({\bf k})}{\omega^2+R^2({\bf k})},
\end{equation}
where $D^\phi_{ij}({\bf k})$ is the pair correlator of the random force
$\phi_i$. It coincides with (\ref{ViVk}) if one choses
(in the momentum representation)
$R({\bf k}) = u_{0} \nu_0 k_{\bot}^{2-\eta}$ and $\phi_{i} = \phi n_{i}$
with $\langle \phi\phi \rangle = g_{0} \nu_0^{3} f_{0}\, \delta(t-t')\,
\delta(k_{\parallel}) k_{\bot}^{5-d-(\varepsilon+\eta)}$.
It remains to note that the resulting velocity ensemble has a finite
correlation time in contrast to the random force $\phi_{i}$ in (\ref{NS}).

\subsection{Feynman diagrammatic technique} \label{sec:DF}

\begin{figure}[h!]
\center
\includegraphics[width=.35\textwidth,clip]{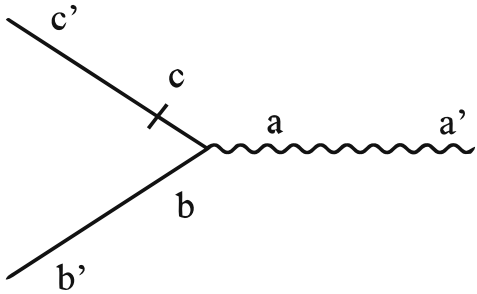}
\caption{The triple vertex with three attached propagators.}
\label{fig:TriVert}
\end{figure}

The model~(\ref{Action}) corresponds to a standard Feynman
diagrammatic technique with the triple vertex
$\theta'\left[-(v_i\partial_i)\theta_k+{\cal A}_0(\theta_i\partial_i)
v_k\right]$
and the three bare propagators. A fragment of arbitrary diagram
is represented in Fig.~(\ref{fig:TriVert}).

In the frequency-momentum representation the triple vertex corresponds
to the expression
\begin{equation}
\label{Triple-vertex}
V_{c\,ab} = i\delta_{bc}\ k_a^{\theta'}
-i{\cal A}_0\delta_{ac}\ k_b^{\theta'},
\end{equation}
where $k^{\theta'}$ is the momentum of the field $\theta'$;
in the diagrams it is represented by the point, in which three lines
connect with each other.
The three propagators are determined by the quadratic
(free) part of the action functional and
are represented in the diagrams as slashed straight (the slashed end
corresponds to the field $\theta'$), straight
(the end without a slash corresponds to the field $\theta$)
and wavy (which corresponds to the field $v$) lines, respectively;
cf.~\cite{VectorN}.

The line $\left\langle v_av_{a'}\right\rangle_0$ in the diagrams corresponds
to the correlation function~(\ref{ViVk}), and the other two
propagators in the frequency-momentum representation have the forms
\begin{eqnarray}
\label{ThThAux-Propgr}
\left\langle \theta_c\theta'_{c'}\right\rangle_0 &=&
\frac{P_{cc'}({\bf k})}{-i\omega+\nu_0\left({\bf k}_
\perp^2+f_0 k_\parallel^2\right)},
\\
\label{ThTh-Propgr}
\left\langle \theta_b\theta_{b'}\right\rangle_0 &=&
\frac{C_{bb'}({\bf k})}{\omega^2+\left[\nu_0\left({\bf k}_\perp^2
+f_0 k_\parallel^2\right)\right]^2}.
\end{eqnarray}
Here $C_{bb'}({\bf k})\propto P_{bb'}({\bf k})$ is the Fourier transform
of the function from (\ref{Cik}); the propagator
$\left\langle \theta'_d\theta'_{d'}\right\rangle$ is equal to zero.

In fact, the action functional (\ref{Action}) has to be modified
for the sake of renormalizability. As a consequence, the functions
(\ref{ThThAux-Propgr}) and~(\ref{ThTh-Propgr}) will acquire certain
additional terms. However, it turns out that those additional
terms do not contribute to the divergent parts of all the relevant
diagrams, and thus they can be neglected. These issues are discussed in
detail in sec.~\ref{sec:TruePropagator}, and in the following we will
use for the propagators
the above expressions (\ref{ThThAux-Propgr}) and~(\ref{ThTh-Propgr}).

%%%%%%%%%%%%%%%%%%%%%%%%%%%%%%%%%%%%%%%%%%%%%%%%%%%%%%%%%%%%%%%%%%%%%%%%%%%%%%%%%%%%%%%%%%%%%%%%%%%%%%%%%%%%%%%%%%%%%%%%%%%%%%%%%%%%%%%%%%%%%%%%%%%%%%%%%%%%%%%%%%%%%%%%
\subsection{Canonical dimensions and UV divergences} \label{sec:Canon}

The analysis of UV divergences is based on the analysis of canonical
dimensions of the 1-irreducible Green functions. In general,
dynamic models have two scales: canonical dimension of some
quantity $F$ (a field or a parameter in the action functional) is completely
characterized by two numbers, the frequency dimension $d_{F}^{\omega}$
and the momentum dimension $d_{F}^{k}$. They are defined such that
\begin{equation}
\label{F-Propto-Dim}
[F] \sim [T]^{-d_{F}^{\omega}} [L]^{-d_{F}^{k}},
\end{equation}
where $L$ is some reference length scale and $T$ is a time scale.

In the {\it scalar} version of strongly anisotropic
model~\eqref{V-n}~--~\eqref{Dv}, however, there are two independent length
scales, related to the directions perpendicular and parallel to the
vector ${\bf n}$~\cite{AntMal2011}. But the transversality conditions
\begin{equation}
\label{transvers}
\partial_i\theta_i=0, \quad \partial_i\theta'_i=0
\end{equation}
forbid this option; see~\cite{VectorN}.
In particular, this means that, in contrast to the scalar case, the
constant $f_0$  from~(\ref{Adv}) in our case is dimensionless.

The dimensions in (\ref{F-Propto-Dim}) are found from the obvious
normalization conditions
$d_{k}^{k}=-d_{\bf x}^{k}=1$,
$d_{k}^{\omega}= -d_{\bf x}^{\omega}=0$,
$d_{\omega}^{\omega} = -d_{t}^{\omega}=1$,
$d_{\omega }^{k}=d_t^{k}=0$,
and from the requirement that each term of the action
functional~(\ref{Action})
be dimensionless (with respect to the two independent dimensions separately).
Based on $d_{F}^{k}$ and $d_{F}^{\omega}$, one can introduce the
total canonical dimension $d_{F}=d_{F}^{k}+2d_{F}^{\omega}$
(in the free theory,
$\partial_{t}\propto\partial^{2}_{\bot} \propto \partial^{2}_{\parallel}$),
which plays in the theory of renormalization of dynamic models the same
role as the conventional (momentum) dimension does in static problems;
see, e.g.,~\cite{Vasiliev-Green}.

\begin{table*}%[b]
\caption{Canonical dimensions of the fields and parameters
%in the model (\protect\ref{Action})
}
\begin{ruledtabular}
%\begin{table}[h]
\label{table1}
\begin{tabular}{c||c|c|c|c|c|c|c|c|c|c|c}
%\hline
$F$ & $\theta' $ & $\theta$ & $ \boldsymbol{v} $ &  $M,m,\mu, \Lambda $ &
$\nu ,\nu_{0}$ & ${\cal A} ,{\cal A}_{0}$ & $f, f_{0}$ & $u, u_0$ & $\alpha_0$
&  $\tilde{g}_{0}$, $g_{0}$ & $\alpha$, $\tilde{g}$, $g$ \\
\hline
%\tableline
$d_{F}^{\omega}$ & 1/2 & $-1/2$ & 1 & 0 & 1 & 0 & 0 & 0 & 0&  0&0\\
\hline
$d_{F}^{k}$ & $d$ & 0 & $-1$ & 1 & $-2$ & 0 & 0 & 0 &$\eta$ & $\xi+\eta$ & 0 \\
\hline
$d_{F}$ & $d+1$ & $-1$ & 1 & 1 & 0 & 0 & 0 & 0 & $\eta$ &
$\xi+\eta$ & 0 \\
%\hline
\end{tabular}
\end{ruledtabular}
\end{table*}

The canonical dimensions for the model~(\ref{Action}) are given in
Table~\ref{table1}, including renormalized parameters, which will be
introduced a bit later. From Table~\ref{table1} it follows that our model
is logarithmic (the coupling constants $g_{0} \sim [L]^{-\xi-\eta}$
and $\alpha_{0} \sim [L]^{-\eta}$
are dimensionless) at $\xi=\eta=0$, so that the UV divergences in the
Green functions manifest
themselves as poles in $\xi$, $\eta$ and their linear combinations.

The total canonical dimension of an arbitrary 1-irreducible Green function
$\Gamma_{N_\Phi} = \langle\Phi \dots \Phi \rangle _{\rm 1-ir}$ is given by
the relation
\begin{equation}
d_{\Gamma_{N_\Phi}}= d+2- \sum_{\Phi} N_{\Phi }d_{\Phi} = d+2-
N_{\theta'} d_{\theta'} - N_{\theta} d_{\theta} - N_{v} d_{v}.
\label{dGamma}
\end{equation}
Here $N_{\Phi}=\{N_{\theta},\,N_{\theta'},\,N_{v}\}$ are the numbers of
corresponding fields entering the function $\Gamma_{N_\Phi}$, and the
summation
over all types of the fields in~(\ref{dGamma}) and analogous formulae below
is always implied.

Superficial UV divergences, whose removal requires counterterms, can be
present only in those functions $\Gamma_{N_\Phi}$ for which the
``formal index of
divergence'' $d_{\Gamma_{N_\Phi}}$ is a non-negative integer.
Dimensional analysis should be augmented by the following considerations:

(1) In any dynamical model of type~(\ref{Action}), 1-irreducible diagrams
with $N_{\theta'}=0$ necessarily contain closed circuits of retarded
propagators~(\ref{ThThAux-Propgr}) or at least one vanishing propagator
$\left\langle \theta_i'\theta_k'\right\rangle$ and therefore vanish.

(2) For any 1-irreducible Green function $N_{\theta'}- N_{\theta}=2N_{0}$,
where $N_{0}\ge0$ is the total number of the bare propagators
$\langle \theta \theta \rangle _0$ entering into any of its diagrams.

(3) Using the transversality condition of the fields $\theta_i$ and $v_i$
we can move one derivative from the vertex
$-\theta'_k(v_i\partial_i)\theta_k+{\cal A}_0\
\theta'_k(\theta_i\partial_i)v_k$
onto the field $\theta'_i$. Therefore, in any
1-irreducible diagram it is always possible to move the derivative onto
external ``tail'' $\theta'_k$, which reduces the real index of divergence:
$d_{\Gamma_{N_\Phi}}' = d_{\Gamma_{N_\Phi}}-N_{\theta'}$.
The field $\theta'_k$ enters the counterterms only in the
form of the derivative $\partial_{i}\theta'_k$.

From Table~\ref{table1} and~(\ref{dGamma}) we find that
\begin{equation}
d_{\Gamma_{N_\Phi}}= d+2 - (d+1) N_{\theta'} + N_{\theta} -N_{v}
\end{equation}
and
\begin{equation}
d'_{\Gamma_{N_\Phi}}\! =(d+2)(1-N_{\theta'}) + N_{\theta} - N_{v}.
\end{equation}
From these expressions we conclude that, for any $d$, superficial
divergences can be present only in the 1-irreducible functions of two types.

The first example is provided by the infinite family of functions
$\langle\theta'\theta\dots\theta\rangle_{\rm 1-ir}$ with $N_{\theta'}=1$
and arbitrary $N_{\theta}$, for which $d_{\Gamma}=2$, $d_{\Gamma}'=0$.
However, all the functions with $N_{\theta}\geq N_{\theta'}$ vanish
(see above) and obviously do not require counterterms. Therefore the
only nonvanishing function from this family is
$\langle\theta_\alpha'\theta_\beta\rangle_{\rm 1-ir}$.

Another possibility is
$\langle \theta'\theta\dots\theta v\dots v\rangle_{\text{1-ir}}$ with
$N_{\theta'}=1$ and arbitrary $N_{\theta}=N_{v}$, for
which $d_{\Gamma}=1$, $d_{\Gamma}'=0$.
From the requirement $N_{\theta}\geq N_{\theta'}$ it follows
that the only nonvanishing function of this type is
$\langle \theta'_\alpha\theta_\beta v_\gamma\rangle_{\text{1-ir}}$.

%%%%%%%%%%%%%%%%%%%%%%%%%%%%%%%%%%%%%%%%%%%%%%%%%%%%%%%%%%%%%%%%%%%%%%%%%%%%%%%%%%%%%%%%%%%%%%%%%%%%%%%%%%%%%%%%%%%%%%%%%%%%%%%%%%%%%%%%%%%%%%%%%%%%%%%%%%%%%%%%%%%%%%%%
\section{Renormalization of the model}\label{sec:Renorm}
%%%%%%%%%%%%%%%%%%%%%%%%%%%%%%%%%%%%%%%%%%%%%%%%%%%%%%%%%%%%%%%%%%%%%%%%%%%%%%%%%%%%%%%%%%%%%%%%%%%%%%%%%%%%%%%%%%%%%%%%%%%%%%%%%%%%%%%%%%%%%%%%%%%%%%%%%%%%%%%%%%%%%%%%

%%%%%%%%%%%%%%%%%%%%%%%%%%%%%%%%%%%%%%%%%%%%%%%%%%%%%%%%%%%%%%%%%%%%%%%%%%%%%%%%%%%%%%%%%%%%%%%%%%%%%%%%%%%%%%%%%%%%%%%%%%%%%%%%%%%%%%%%%%%%%%%%%%%%%%%%%%%%%%%%%%%%%%%%
\subsection{Perturbation expansion for the 1-irreducible linear response
function}

The field theoretic formulation means that statistical averages of random
quantities in the stochastic problem~(\ref{Adv}),~(\ref{ViVk})
coincide with functional averages with weight $\exp {\cal S}(\Phi)$
with the action~(\ref{Action}).

Let us denote the generating functional of the normalized full Green
functions $G = \langle\Phi \dots \Phi \rangle$ as $G(\widetilde{A})$, where
$\widetilde{A}(x)=\left\{A(x), A'(x), A_v(x)\right\}$ is the set of
``sources,'' arbitrary functional arguments of the same nature as
the corresponding fields.
Thus, the generating functional of the 1-irreducible Green
functions is obtained using the Legendre transform:
\begin{equation}
\label{Legendre}
\Gamma\left(\Phi\right)=\ln G(\widetilde{A})-\Phi\widetilde{A};
\end{equation}
see, e.g.,~\cite{Vasiliev-Green}.

The Green functions with the auxiliary field $\boldsymbol{\theta}'$
represent, in the field theoretic formulation, the response functions of
the original stochastic problem, in particular,
the simplest (linear) response function is given by the relation
\begin{equation}
%\label{Delta-Green}
\left\langle \delta\theta_\beta/\delta f_\alpha\right\rangle
= \left\langle\theta_\beta \theta'_\alpha\right\rangle.
\end{equation}

\begin{figure}[h!]
\center
\includegraphics[width=.35\textwidth,clip]{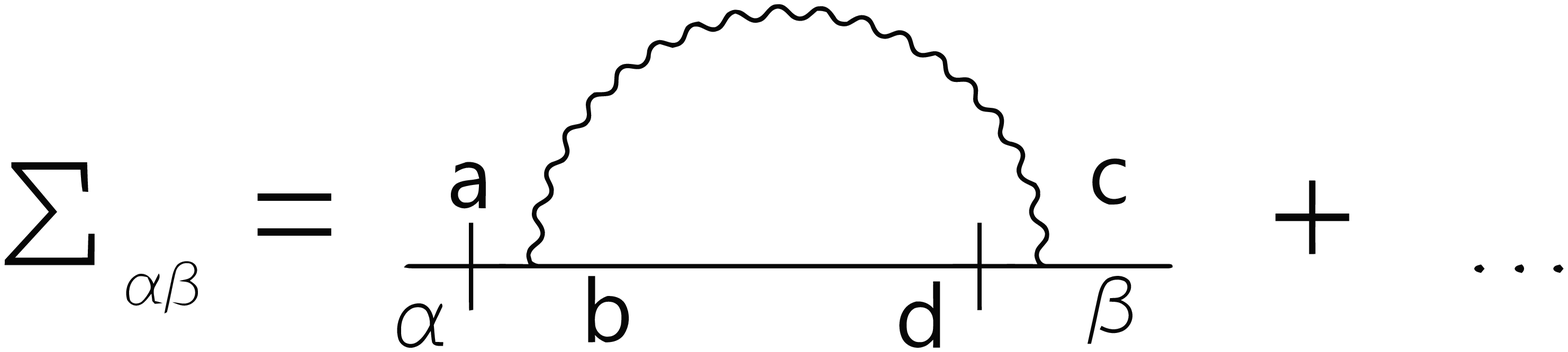}
\caption{Diagrammatic representation for $\Sigma_{\alpha\beta}$.}
\label{fig:SelfEnergy}
\end{figure}

Let us consider the 1-irreducible linear response function
\begin{equation}
\left.\Gamma_2^{\alpha\beta}=\langle\theta_\alpha'\theta_\beta
\rangle_{\rm 1-ir}=
\frac{\delta}{\delta \theta'_\alpha}\frac{\delta}{\delta \theta_\beta}
\Gamma(\Phi)\right|_{\Phi=0}.
\label{Vasik2}
\end{equation}
In accordance with~\eqref{Legendre} generating function for it
%of the 1-irreducible Green functions $\Gamma(\Phi)$
consists of two parts,
\begin{equation}
\Gamma(\Phi) =  {\cal S}(\Phi) + \widetilde{\Gamma}(\Phi),
\end{equation}
where for the functional arguments we have used the same symbols
$\Phi=\left\{\boldsymbol{\theta},\boldsymbol{\theta}',
\boldsymbol{v}\right\}$ as for the corresponding random fields;
${\cal S}(\Phi)$ is the action functional (\ref{Action}) and
$\widetilde{\Gamma}(\Phi)$ is the sum of all the 1-irreducible diagrams
with loops.
Thus, %for the function (\ref{Vasik2})
one obtains
\begin{widetext}
\begin{equation}
\label{Dyson}
\Gamma_2^{\alpha\beta}=i\omega P_{\alpha\beta}({\bf p})-\nu_0
{\bf p}_\perp^2 P_{\alpha\beta}({\bf p})-\nu_0f_0({\bf pn})^2
P_{\alpha\beta}({\bf p})+\Sigma_{\alpha\beta},
\end{equation}
\end{widetext}
where $P_{\alpha\beta}({\bf p})=\delta_{\alpha\beta}-p_\alpha p_\beta/p^2$
is transverse projector and $\Sigma_{\alpha\beta}$ is the
``self-energy operator,''
diagrammatic representation for which is represented in the
Fig.~\ref{fig:SelfEnergy}.
\noindent
Here the ellipsis stands for the 2-, 3- and other N-loop diagrams.

The typical feature of all rapid-change models like~(\ref{Kraich}) with
retarded bare propagator of the type~(\ref{ThThAux-Propgr})
is that all the skeleton multiloop diagrams entering into the self-energy
operator contain closed circuits of such retarded propagators
and therefore vanish~\cite{RG,AntGul2012-mod,VectorN}. The dependence of
the frequency in function $D_v$ [see~\eqref{Dv}] destroys this easy
construction, and now all the N-loop diagrams are expected to give some
nontrivial contribution to the function $\Sigma_{\alpha\beta}$.

Let us start with the one-loop diagram. It is represented by the expression
\begin{widetext}
\begin{equation}
\label{Sigma-Expr}
\Sigma_{\alpha\beta}=D_0\int\frac{d\omega}{2\pi}\int\frac{d{\bf k}}
{(2\pi)^d}\frac{2\pi\, \delta(k_\parallel)k_\perp^{5-d-(\xi+\eta)}}
{\left(-i\omega+\nu_0\left[\left({\bf p+k}\right)_\perp^2+f_0
\left(p+k\right)_\parallel^2\right]\right)\left(\omega^2+\left[\alpha_0
\nu_0k_\perp^{2-\eta}\right]^2\right)}
P_{\alpha i}({\bf p}) J_{ij} P_{j\beta}({\bf p}),
\end{equation}
\end{widetext}
where the fraction is a product of the propagator
function~\eqref{ThThAux-Propgr} and the correlator~\eqref{Dv},
transverse projectors $P_{\alpha i}({\bf p})$ and $P_{\beta j}({\bf p})$
are present due to the transversality conditions~\eqref{transvers},
and $J_{ij}$ is an index structure of this diagram:
\begin{equation}
J_{ij}= V_{i\, ab}({\bf p})V_{d\,cj}({\bf p+k})P_{bd}({\bf p+k})n_an_c.
\end{equation}
Here and below $V_{ijk}({\bf p})$ is the triple vertex~(\ref{Triple-vertex});
the Greek letters $\alpha$, $\beta$ and the Roman letters $a$--$d$
denote the vector indices of the propagators~(\ref{ViVk})
and~(\ref{ThThAux-Propgr}) with the implied summation
over repeated indices. Since the index of divergence for this diagram
$d_\Gamma=2$, we need to calculate only the terms, proportional
to ${\bf p}^2$.

The calculation of this diagram is similar to the zero-time correlation
case~\cite{VectorN}, so we will discuss it here only briefly.

The integration over the frequency $\omega$ is trivial. In order to
integrate over the  vector ${\bf k}$ with the function $\delta(k_\parallel)$
in the integrand we need to
average the expression~\eqref{Sigma-Expr} over the angles:
\begin{equation}
\label{Int-Aver-Angles}
\int d{\bf k}\,\delta(k_\parallel) f({\bf k})=
S_{d-1} \int_m^\infty dk_{\perp}\, k_{\perp}^{d-2}\,
\left\langle f({\bf k}_{\perp})\right\rangle,
\end{equation}
where $\langle\cdots\rangle$ is the averaging over the unit sphere in the
$(d-1)$-dimensional space, $S_{d-1}$ is its surface area, and
$k_{\perp}= |{\bf k}_{\perp}|$. To average some function of $k_\perp$ over
the angles in the orthogonal subspace we use the following expression:
\begin{equation}
\label{k-transv-Aver-Angles}
\left\langle \frac{k_i^\perp k_j^\perp}{k_\perp^2} \right\rangle
= \frac{P_{ij}({\bf n})}{(d-1)}.
\end{equation}
This gives:
\begin{widetext}
\begin{equation}
\label{Sigma-eta}
\Sigma_{\alpha\beta} = -\frac{g_0\nu_0f_0}{2\alpha_0} C_{d-1}
\left[\frac{d-2+{\cal A}_0}{d-1}
P_{\alpha\beta}({\bf p})+\frac{({\cal A}_0-1)^2}{d-1}
\hat{n}_\alpha \hat{n}_\beta\right]({\bf pn})^2
\int_m^\infty dk_\perp\frac{ k_\perp^{1-\xi}}
{k_\perp^2+\alpha_0k_\perp^{2-\eta}},
\end{equation}
\end{widetext}
where $C_{d-1}\equiv S_{d-1}/(2\pi)^{d-1}$ and the vector $\hat{n}_k$,
which is orthogonal to ${\bf p}$, is defined as
\begin{equation}
\hat{n}_k = P_{mk}({\bf p})n_m = n_k-p_\parallel p_k/p^2.
\end{equation}

The integral over $k_\perp$ in expression~\eqref{Sigma-eta} can be
simplified in the minimal subtraction (MS) renormalization scheme,
which we adopt in what follows. In that scheme, all the anomalous
dimensions $\gamma$ are independent of the regularizators like $\xi$
and $\eta$, and we may chose them arbitrary with the only restriction~--
our diagrams have to remain UV finite; see~\cite{FinTimeEta} for
detailed discussion. The most convenient way is to put $\eta=0$,
so the expression~\eqref{Sigma-eta} turns into
\begin{widetext}
\begin{equation}
%\label{Sigma-eta}
\Sigma_{\alpha\beta} = -\frac{g_0\nu_0f_0}{2\alpha_0(1+\alpha_0)} C_{d-1}
\left[\frac{d-2+{\cal A}_0}{d-1}
P_{\alpha\beta}({\bf p})+\frac{({\cal A}_0-1)^2}{d-1}
\hat{n}_\alpha \hat{n}_\beta\right]({\bf pn})^2
\int_m^\infty dk_\perp\frac{ 1}{k_\perp^{1+\xi}}
\end{equation}
%\end{widetext}
and we obtain the following result:
%\begin{widetext}
\begin{equation}
\label{Sigma-Answer}
\Sigma_{\alpha\beta} = -\frac{g_0\nu_0f_0}{2\alpha_0(1+\alpha_0)} C_{d-1}
\left[\frac{d-2+{\cal A}_0}{d-1}
P_{\alpha\beta}({\bf p})+\frac{({\cal A}_0-1)^2}{d-1}
\hat{n}_\alpha \hat{n}_\beta\right]({\bf p\cdot n})^2
\frac{m^{-\xi}}{\xi}.
\end{equation}
\end{widetext}

The remaining multiloop diagrams will be discussed a bit later,
in section~\ref{sec:Multiloop}.

%%%%%%%%%%%%%%%%%%%%%%%%%%%%%%%%%%%%%%%%%%%%%%%%%%%%%%%%%%%%%%%%%%%%%%%%%%%%%%%%%%%%%%%%%%%%%%%%%%%%%%%%%%%%%%%%%%%%%%%%%%%%%%%%%%%%%%%%%%%%%%%%%%%%%%%%%%%%%%%%%%%%%%%%
\subsection{Perturbation expansion for the 1-irreducible function
$\langle \theta'_\alpha\theta_\beta v_\gamma\rangle_{\text{1-ir}}$}

\begin{figure}[h!]
\center
\includegraphics[width=.25\textwidth,clip]{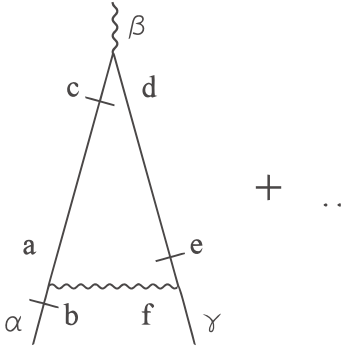}
\caption{Diagrammatic representation for $\Delta_{\alpha\,\beta\gamma}$.}
\label{fig:Delta}
\end{figure}

The expansion like~\eqref{Dyson} for the function
$\langle \theta'_\alpha\theta_\beta v_\gamma\rangle_{\text{1-ir}}$
has the form
\begin{eqnarray}
\langle \theta'_\alpha\theta_\beta v_\gamma\rangle_{\text{1-ir}} &=&
V_{\alpha\,\beta\gamma} + \Delta_{\alpha\,\beta\gamma} \nonumber \\
&=& i\delta_{\alpha\gamma}p_\beta-i{\cal A}_0\delta_{\alpha\beta}p_\gamma
+ \Delta_{\alpha\,\beta\gamma},
\end{eqnarray}
where $V_{\alpha\,\beta\gamma}$ is the vertex~\eqref{Triple-vertex} and
$\Delta_{\alpha\,\beta\gamma}$ is represented in the Fig.~\ref{fig:Delta}.

As in the case of self energy operator in Fig.~\ref{fig:SelfEnergy},
the ellipsis stands for the 2-, 3- and other N-loop diagrams.

Since our model is Galilean invariant, as discussed in Sec.~\ref{sec:Model},
the terms $\theta'_k\partial_t\theta_k$ and
$\theta'_k(v_i\partial_i)\theta_k$ in the action functional may be
renormalized only with the only renormalization constant $Z_1$. The index
of divergence for this function is $d_\Gamma=1$, so that the counterterms
with $\partial_t$ are forbidden. Consequently, counterterm
$\theta'_k(v_i\partial_i)\theta_k$ is also forbidden. If ${\cal A}_0=1$,
the vertex~\eqref{Triple-vertex} is transverse, the nonlocal term
$\partial {\cal P}$ in the stochastic equation~\eqref{Adv} is absent
and the action functional is local in-time. This means that the
counterterm $\theta'_k(\theta_i\partial_i)v_k$ is forbidden because
the appearance of some constant $Z_2$, which this term is renormalized by,
is equivalent to appearance of some multiplier like ${\cal A}_0\neq1$,
i.e., the appearance of nonlocal terms in the action functional.
Similar reasoning exclude the appearance of such a counterterm
if ${\cal A}_0=0$. Thus, we may conclude, that
$\Delta_{\alpha\beta\gamma}$ is proportional to ${\cal A}_0({\cal A}_0-1)$
and vanish for the aforementioned cases.

The procedure of calculating the one-loop approximation of
$\Delta_{\alpha\,\beta\gamma}$ is similar to the one-loop
contribution to the self-energy operator $\Sigma_{\alpha\beta}$,
discussed in previous section. The analytical expression for the former is
\begin{widetext}
\begin{eqnarray}
\label{Delta-Expr}
\Delta_{\alpha\beta\gamma}=D_0\int\frac{d\omega}{2\pi}\int\frac{d{\bf k}}
{(2\pi)^d}\frac{1}
{\left(-i\omega+\nu_0\left[({\bf k+q})_\perp^2+f_0(k+q)_\parallel^2\right]
\right)
\left(-i\omega+\nu_0\left[({\bf k-p})_\perp^2+f_0(k-p)_\parallel^2\right]
\right)}
\noindent \nonumber \\
\times\frac{2\pi\, \delta(k_\parallel)k_\perp^{5-d-(\xi+\eta)}}
{\left(\omega^2+\left[\alpha_0\nu_0k_\perp^{2-\eta}\right]^2\right)}
P_{\alpha i}({\bf q}) J_{i\beta j} P_{j\gamma}({\bf p})n_\beta,
\end{eqnarray}
\end{widetext}
where ${\bf p}$ and ${\bf q}$ are two external momenta, $J_{i\beta j}$ is
the index structure of this diagram, transverse projectors
$P_{i\alpha}({\bf p})$ and $P_{j\gamma}({\bf p})$ and vector $n_\beta$
are present due to the transversality conditions~\eqref{transvers} and
definition~\eqref{V-n}. Since the index of divergence for this function
is $d_\Gamma=1$, we need to calculate only the term, proportional to the
linear combination of ${\bf p}$ and ${\bf q}$. Also we may put $\eta=0$
in this diagram and left with the only regularizator $\xi$.

The integral over $\omega$ is convergent; direct calculation shows that
\begin{equation}
J_{i\beta j} \propto  {\cal A}_0(1-{\cal A}_0)
n_iP_{\beta j}({\bf n}).
\end{equation}
This means that
\begin{equation}
\label{Delta-answer}
J_{\alpha\beta\gamma} \equiv P_{\alpha i}({\bf q}) J_{i\beta j}
P_{j\gamma}({\bf p})n_\beta =0,
\end{equation}
i.e., the function
$\langle \theta'_\alpha\theta_\beta v_\gamma\rangle_{\text{1-ir}}$
does not diverge not only for the  cases ${\cal A}_0=0$
and ${\cal A}_0=1$, discussed above, but also in all the other situations.

The multiloop diagrams will be discussed in the next subsection.

%%%%%%%%%%%%%%%%%%%%%%%%%%%%%%%%%%%%%%%%%%%%%%%%%%%%%%%%%%%%%%%%%%%%%%%%%%%%%%%%%%%%%%%%%%%%%%%%%%%%%%%%%%%%%%%%%%%%%%%%%%%%%%%%%%%%%%%%%%%%%%%%%%%%%%%%%%%%%%%%%%%%%%%%
\subsection{Multiloop diagrams}
\label{sec:Multiloop}

\begin{figure}[h]
\center
\includegraphics[width=.15\textwidth,clip]{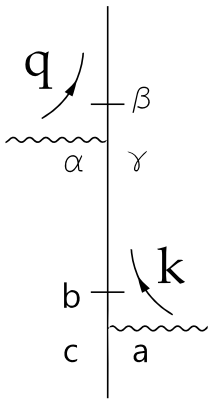}
\caption{Fragment of {\it arbitrary} multiloop diagram, entering into expansion of the function $\langle \theta'_\alpha\theta_\beta v_\gamma\rangle_{\text{1-ir}}$.}
\label{fig:Multi-Loop}
\end{figure}

In order to renormalize our model we have to deal with two types
of multiloop diagrams~-- one of types corresponds to the
function $\langle\theta_\alpha'\theta_\beta
\rangle_{\rm 1-ir}$ and is represented in Fig.~\ref{fig:SelfEnergy},
the other one corresponds to the function
$\langle \theta'_\alpha\theta_\beta v_\gamma\rangle_{\text{1-ir}}$
and is represented in Fig.~\ref{fig:Delta}. Let us start with the latter.
Any multiloop diagram of this type contains a part with the structure,
represented in Fig.~\ref{fig:Multi-Loop}.
\noindent
Since it is sufficient to calculate all the diagrams at external momenta
equal to zero (the real index of divergence $d_{\Gamma}'=0$), the integral,
corresponding to the divergent part of
the diagram, necessarily contains as a factor the following expression:
\begin{equation}
I_0\propto\delta(k_\parallel)\delta(q_\parallel)n_aV_{bac}({\bf k})
n_\alpha V_{\beta\alpha\gamma}({\bf k+q})P_{\gamma b}({\bf k}),
\end{equation}
where $V$ is the vertex~(\ref{Triple-vertex}), and the $\delta$-functions
appear from velocity correlator~(\ref{ViVk}). Since $I_0$ is proportional
to the sum of $k_\parallel$ and $q_\parallel$ with some coefficients, after
integration with the $\delta$-functions all these diagrams vanish.

\begin{figure}[h!]
\center
\includegraphics[width=.22\textwidth,clip]{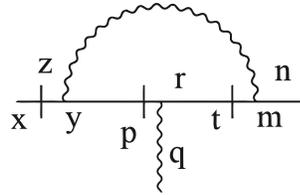}
\caption{One of two possible fragments of {\it arbitrary} multiloop
diagram for self-energy operator $\Sigma_{\alpha\beta}$.}
\label{fig:Multi-Loop-Delta-1}
\end{figure}

\begin{figure}[h!]
\center
\includegraphics[width=.25\textwidth,clip]{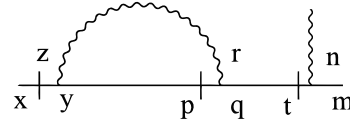}
\caption{Another possible fragment of {\it arbitrary} multiloop diagram for self-energy operator $\Sigma_{\alpha\beta}$.}
\label{fig:Multi-Loop-Delta-2}
\end{figure}

Any multilop diagram, entering into the expansion of the 1-irreducible
linear response
function $\langle\theta_\alpha'\theta_\beta\rangle_{\rm 1-ir}$, contains
a part with structure,
represented in Fig.~\ref{fig:Multi-Loop-Delta-1} or a part with structure,
represented in Fig.~\ref{fig:Multi-Loop-Delta-2}.

Since in any
1-irreducible diagram it is always possible to move the derivative onto
external ``tail'' $\theta'_k$, the
real index of divergence for this diagram $d_{\Gamma}' = 1$.
This means, that in course of calculation of the structures, represented
in Fig.~\ref{fig:Multi-Loop-Delta-1} and Fig.~\ref{fig:Multi-Loop-Delta-2},
we are interested only in terms, linear in the external momenta ${\bf p}$.

The analytical expression for the first structure, denoted by $I_1$,
is proportional to
\begin{eqnarray}
\label{MLoop}
I_1 &\propto & \delta(k_\parallel)\delta(q_\parallel)n_z
V_{xyz}({\bf p+k})P_{yp}({\bf p+k-q})
 \\
& \times & n_qV_{pqr}({\bf p+k-q})P_{rt}({\bf p-q})n_nV_{tnm}({\bf p-q}).
\nonumber
\end{eqnarray}
Here ${\bf p}$ is the external momentum,
${\bf k}$ and ${\bf q}$ are internal integration momenta,
$V$ is the vertex~(\ref{Triple-vertex}), $P$ is the transverse projector,
and the unit vector ${\bf n}$ and $\delta$-functions
stem from velocity correlator~(\ref{ViVk}).

Direct calculation shows, that $I_1$ is proportional to some linear
combination of $k_\parallel$ and $q_\parallel$, and, as well as in
the case of $I_0$, after the integration with the $\delta$-functions all
diagrams with this structure vanish.

Another structure, represented in Fig.~\ref{fig:Multi-Loop-Delta-2},
possess the same property~-- analytical expression for it is
similar to~\eqref{MLoop}, and, as can be seen from the direct calculation,
all the diagrams with this structure also appear to be equal to zero.

It should be stressed that, in contrast to rapid-change models
like~\eqref{Kraich} with $\delta$ functions in time, where all
these multiloop diagrams vanish
due to the closed circuits of retarded propagators, in our model
their vanishing has a rather nontrivial origin and results
from the presence of the anisotropy in it.

%%%%%%%%%%%%%%%%%%%%%%%%%%%%%%%%%%%%%%%%%%%%%%%%%%%%%%%%%%%%%%%%%%%%%%%%%%%%%%%%%%%%%%%%%%%%%%%%%%%%%%%%%%%%%%%%%%%%%%%%%%%%%%%%%%%%%%%%%%%%%%%%%%%%%%%%%%%%%%%%%%%%%%%%
\subsection{Renormalization and RG equations}
\label{sec:TruePropagator}

Substitution of the explicit expression~(\ref{Sigma-Answer}) for the
divergent part of the self-energy operator $\Sigma_{\alpha\beta}$
into the expression~(\ref{Dyson}) for the 1-irreducible linear response
function $\Gamma_2^{\alpha\beta}$ gives
\begin{widetext}
\begin{equation}
\label{Dyson-2}
\Gamma_2^{\alpha\beta}= \{ i\omega -\nu_0 {\bf p}_\perp^2
-\nu_0f_0({\bf p\cdot n})^2 \} \ P_{\alpha\beta}({\bf p})
-\frac{g_0\nu_0f_0}{2\alpha_0(1+\alpha_0)}
\left[\frac{(d-2+{\cal A}_0)}{d-1}
 P_{\alpha\beta}({\bf p})+\frac{({\cal A}_0-1)^2}{d-1}
\hat{n}_\alpha \hat{n}_\beta\right] C_{d-1}({\bf p\cdot n})^2\times
\frac{m^{-\xi}}{\xi}.
\end{equation}
\end{widetext}

The renormalization constants are found from the requirement that the
function (\ref{Dyson-2}), when expressed in new renormalized variables,
be UV finite, i.e., finite at $\xi\to0$. From the analysis of this
expression it follows, however, that the pole in $\xi$ in the structure
with $\hat{n}_\alpha \hat{n}_\beta$ cannot be removed by renormalization
of the model parameters because the bare part of $\Gamma_2^{\alpha\beta}$
does not contain analogous term. In order to ensure multiplicative
renormalizability one has to add such term, with a new positive amplitude
factor $u_{0}$, to the bare part:
\begin{widetext}
\begin{eqnarray}
\Gamma_2^{\alpha\beta} &=& \left\{i\omega
-\nu_0{\bf p}_\perp^2
-\nu_0f_0({\bf p\cdot n})^2\right\}\ P_{\alpha\beta}({\bf p})
-\nu_0f_0u_0\ ({\bf p\cdot n})^2
\hat{n}_\alpha \hat{n}_\beta
\nonumber \\
&-& \frac{g_0\nu_0f_0}{2\alpha_0(1+\alpha_0)} \left[\frac{(d-2+{\cal A}_0)}{d-1}
P_{\alpha\beta}({\bf p})+\frac{({\cal A}_0-1)^2}{d-1}
\hat{n}_\alpha \hat{n}_\beta\right] C_{d-1}({\bf p\cdot n})^2\times
\frac{m^{-\xi}}{\xi}.
\label{Dyson-True}
\end{eqnarray}
\end{widetext}
This means that the original model (\ref{Action}) is extended by adding
a new term of the form
$u_{0}f_{0}\nu_0(n_k\theta_k') \partial_\parallel^2(n_k\theta_k)$;
the interpretation of the new parameter $u_{0}$ is literally the same as
for $f_{0}$ in Sec.~\ref{sec:Model}.

Now the model is multiplicatively renormalizable with two independent
renormalization constants $Z_{f}$ and $Z_{u}$:
\begin{eqnarray}
\label{Renorm-Parameters}
\nu_0=\nu Z_{\nu}, \ f_0=f Z_{f}, \ u_0=u Z_{u},\  \nonumber \\
{\cal A}_0={\cal A}Z_{\cal A},\ g_{0}=g\mu^{\xi+\eta}Z_{g}, \
\alpha_{0}=\alpha\mu^{\eta}Z_{\alpha},
\end{eqnarray}
at that
\begin{equation}
\label{Renorm-Parameters-1}
Z_\nu=Z_\alpha=Z_{\cal A}=1,\ Z_{g}=Z_{f}^{-1}.
\end{equation}
Here $\mu$ is the ``reference mass'' (additional free parameter of the
renormalized theory) in the MS renormalization
scheme, which we always use in what follows;
$g$, $u$, $\alpha$, $\nu$, ${\cal A}$ and $f$
are renormalized analogs of the bare parameters
$g_{0}$, $u_0$, $\alpha_0$, $\nu_0$, ${\cal A}_0$ and $f_0$,
and $Z_i=Z_i(g,\xi,d)$ are the renormalization constants. Their
relations in~(\ref{Renorm-Parameters-1}) result from the absence of
renormalization of the contribution with $D_v^{-1}$ in~(\ref{Action}),
so that $D_{0}\equiv g_{0}\nu_0^3f_0 = g\mu^{\xi+\eta} \nu^3 f$,
$\alpha_0\nu_0=\alpha\mu^\eta\nu$.
No renormalization of the fields and the parameter
$m_{0}=m$ is needed: i.e., $Z_{\Phi}=1$ for all $\Phi$ and $Z_{m}=1$.

The renormalized action functional has the form
\begin{widetext}
\begin{eqnarray}
\label{RenAction}
{\cal S}_R(\Phi) &=& \frac{1}{2}\theta'_iD_{\theta}\theta'_k
-\frac{1}{2}v_iD_v^{-1}v_k+
\theta'_k\left[-\partial_t\theta_k-(v_i\partial_i)\theta_k+{\cal A}
(\theta_i\partial_i)v_k+\nu(\partial_\perp^2+fZ_f \partial_\parallel^2)
\theta_k\right]+
\nonumber \\
&+& \nu\, fZ_f\, uZ_u\, (n_k\theta_k')
\partial_\parallel^2(n_k\theta_k),
\end{eqnarray}
\end{widetext}
where the function $D_v$ from~(\ref{Dv}) should be expressed in renormalized
variables using~(\ref{Renorm-Parameters}).

At this moment one important problem springs up.
Since the original model is extended by introducing a new term
(proportional to the $\theta'_i\theta_k$)
in the action functional~\eqref{Action},
one may guess that the propagator functions~\eqref{ThThAux-Propgr}
and~\eqref{ThTh-Propgr} have to be modified. Consequently, we have to
recalculate the diagrams for functions
$\langle\theta_\alpha'\theta_\beta \rangle_{\rm 1-ir}$ and
$\langle \theta'_\alpha\theta_\beta v_\gamma\rangle_{\text{1-ir}}$, i.e.,
the expressions~\eqref{Sigma-Answer} and~\eqref{Delta-answer}.

If fact, the difference between the original expressions for the bare
propagators and the new ones is that the second have additional terms,
which are proportional to the $p_\parallel$. Consequently, they do not
contribute to the integrals and revision of the
final expressions is in fact not needed; this means that
expressions~\eqref{Sigma-Answer} and~\eqref{Delta-answer} remain
valid in the modified model. This problem was examined in details
in~\cite{VectorN}; moreover, the derivation of the propagators in
the presence of a distinguished direction ${\bf n}$, i.e., in fact,
the matrix inversion in the orthogonal subspace,
was also discussed there.

%\subsection{RG equations and fixed points} \label{sec:FP}

Now we are ready to study the fixed points $\left\{g_i^*\right\}$
that govern the IR asymptotic behavior. The basic RG equation for a
multiplicatively renormalizable quantity (correlation function,
composite operator, etc.) has the form
\begin{equation}
\label{RG-Eqtn}
\bigl[{\cal D}_{RG}+ \gamma_{F} \bigr] F_{R}=0
\end{equation}
and is a consequence of operating on the relation $F=Z_{F}F_{R}$ with the
differential operation $\mu\partial_{\mu}$ for fixed set of bare parameters
$e_{0}$ $=$ $\left\{g_{0}, \nu_{0}, f_{0}, u_{0}, {\cal A}_{0}\right\}$.
This operation is customarily denoted as $\widetilde{\cal D}_\mu$, and
$\gamma_F$ is the anomalous dimension of $F$. Since $Z_\nu=1$, the
renormalization group operator ${\cal D}_{RG}$ has the form
${\cal D}_{RG} = {\cal D}_{\mu} + \beta_g \partial_{g} -
\gamma_{f}{\cal D}_{f}- \gamma_{u}{\cal D}_{u}$, where
${\cal D}_{x}\equiv x\partial_{x}$ for any variable $x$.

The RG functions are defined as
\begin{subequations}
\begin{equation}
\label{Beta-g-Def}
\beta_g \equiv \widetilde{\cal D}_\mu g =g\,[-\xi-\eta-\gamma_{g}(g)],
\end{equation}
\begin{equation}
\label{Beta-u-Def}
\beta_u \equiv \widetilde{\cal D}_\mu u =-u\gamma_{u}(g,u),
\end{equation}
\begin{equation}
\label{Beta-alpha-Def}
\beta_\alpha \equiv \widetilde{\cal D}_\mu \alpha =-\eta\alpha,
\end{equation}
\begin{equation}
\gamma_{F} \equiv \widetilde{\cal D}_\mu \ln Z_{F} =
\beta_g \partial_{g} \ln Z_{F} \quad  {\rm for\ any\ } Z_{F}.
\end{equation}
\end{subequations}
The relations between $\beta$ and $\gamma$
in~(\ref{Beta-g-Def})~--~\eqref{Beta-alpha-Def}
result from their definitions along with relations~(\ref{Renorm-Parameters})
and~\eqref{Renorm-Parameters-1}.

The constants $Z_i$ are found from the requirement of UV finiteness of the
expression~(\ref{Dyson-True}). Thus, for
the parameter $f_0$ that splits the Laplace operator we obtain
\begin{equation}
Z_f=1-\frac{(d-2+{\cal A})}{2(d-1)}\
\frac{g}{\alpha(\alpha+1)}\frac{1}{\xi}+O\left(g^2\right),
\end{equation}
\begin{equation}
\label{gamma-f}
\gamma_f=\frac{\left(d-2+{\cal A}\right)}{2(d-1)}\
\frac{g}{\alpha(\alpha+1)},
\end{equation}
where we passed to the new coupling constant $g\equiv\tilde{g}\,C_{d-1}$
with $C_{d-1}$ from (\ref{Sigma-Answer}).

Then we have to renormalize the constant $u_0$ such that the expression
\begin{equation}
g_0f_0u_0\left[1+\frac{({\cal A}-1)^2}{2(d-1)}
\frac{1}{u_0\,\alpha_0(1+\alpha_0)}\times\frac{m^{-\xi}}{\xi}\right]
n_\alpha n_\beta({\bf p\cdot n})^2
\end{equation}
be UV finite to the first order in $g$. Therefore,
\begin{equation}
Z_uZ_f=1-\frac{({\cal A}-1)^2}
{2(d-1)}\ \frac{g}{u\,\alpha(1+\alpha)}\ \frac{1}{\xi}+O\left(g^2\right),
\end{equation}
and
\begin{equation}
\gamma_u+\gamma_f=\frac{({\cal A}-1)^2}{2(d-1)}\
\frac{g}{u\,\alpha(1+\alpha)},
\end{equation}
where the constant $\gamma_f$ is obtained in~\eqref{gamma-f}.
Furthermore, from the last relation in~(\ref{Renorm-Parameters-1})
it follows that for the coupling constant $g$
\begin{equation}
\label{gamma-g-gamma-f}
\gamma_g=-\gamma_f= -\frac{(d-2+{\cal A})}{2(d-1)}\
\frac{g}{\alpha(1+\alpha)}.
\end{equation}

We stress that, since the expression~(\ref{Dyson-True}) is exact, i.e.,
it has no corrections in coupling constant $g$, all the above expressions
for the anomalous dimensions $\gamma_{f,g,u}$ are exact, too.

%%%%%%%%%%%%%%%%%%%%%%%%%%%%%%%%%%%%%%%%%%%%%%%%%%%%%%%%%%%%%%%%%%%%%%%%%%%%%%%%%%%%%%%%%%%%%%%%%%%%%%%%%%%%%%%%%%%%%%%%%%%%%%%%%%%%%%%%%%%%%%%%%%%%%%%%%%%%%%%%%%%%%%%%
\subsection{Fixed points}

One of the basic RG statements is that the asymptotic behavior of the model
is governed by the fixed points $\left\{g^*, \alpha^*,u^*, f^*\right\}$,
defined by the relations
\begin{equation}
\label{Beta-zero}
\beta_g=0,\quad\beta_u=0, \quad \beta_f=0\quad\text{and}\quad
\beta_\alpha=0;
\end{equation}
here
\begin{subequations}
\label{Beta}
\begin{equation}
\beta_g=g\left(-\xi-\eta+\gamma_f\right)=g\left[-\xi-\eta+
\frac{(d-2+{\cal A})}{2(d-1)}\ \frac{g}{\alpha(1+\alpha)}\right],
\end{equation}
\begin{equation}
\beta_u=-u\gamma_u=\frac{g}{\alpha(\alpha+1)}
\left[\frac{(d-2+{\cal A})}{2(d-1)}u-\frac{({\cal A}-1)^2}{2(d-1)}\right],
\end{equation}
\begin{equation}
\beta_f=-f\gamma_f=-f\frac{(d-2+{\cal A})}{2(d-1)}\
\frac{g}{\alpha(1+\alpha)},
\end{equation}
\end{subequations}
the expression for $\beta_\alpha$ is written in~\eqref{Beta-alpha-Def}.

The type of a fixed point (IR/UV attractive or a saddle point), i.e., the
character of the RG flow in vicinity of the point, is determined by the
matrix $\Omega_{ik} = \partial\beta_i/\partial g_k$, where $\beta_i$ is
the full set of $\beta$-functions and $g_k$ is the full set of couplings.
For an IR attractive fixed point the matrix $\Omega$ are positive,
i.e., the real parts of all its eigenvalues are positive.

The analysis of the $\beta$-functions reveals several fixed points.
The first possibility is to put $\alpha^*=0$; consequently we get at once
the trivial case $g^*=0$. There is, however, another possibility~-- to
disclose it we have to pass from the coupling constant $g$ to new constant
$g'=g/\alpha$, which is assumed to be finite at $\alpha\to0$. In fact this
means, that the correlation function $D_v(\omega)$ becomes proportional to
$\delta(\omega)$ (see~\eqref{Dv}) and we deal with the independent of time
(``frozen'' or ``quenched'') velocity field.

The new $\beta$-function, which remains nonzero at $\alpha\to0$, is
\begin{equation}
\beta_{g'}=\frac{1}{\alpha}\beta_{g}-\frac{g}{\alpha^2}
\beta_{\alpha}=g'\left[-\xi+\frac{(d-2+{\cal A})}{2(d-1)}g'\right];
\end{equation}
the matrix $\Omega$ in these variables has the form
\begin{equation}
\label{Omega}
\Omega =
\begin{pmatrix}
\partial_{g'}\beta_{g'} & \partial_{g'}\beta_u & 0& \partial_{g'}\beta_f \\
0 & \partial_{u}\beta_{u} & 0 & 0  \\
0 & 0 & \partial_{\alpha}\beta_{\alpha} & 0 \\
0 & 0 & 0  &  \partial_{f}\beta_{f}\\
\end{pmatrix}.
\end{equation}
This situation implies two options:

(1a) $g'^*=0$, with
$\Omega_{g'g'}^*=\left.\partial\beta_{g'}/\partial g'\right|_{g'=g'^*}=-\xi$
and
$\Omega_{\alpha\alpha}^*=-\eta$.

For the two remaining parameters $u$ and $f$ we have
$\beta_u=\beta_f\equiv0$, $\Omega_{uu}^*=\Omega_{ff}^*\equiv0$,
so that both $u$ and $f$ remain free parameters.

Since $\Omega_{g'u}^*=0$, the matrix $\Omega$ is triangle and its
eigenvalues coincide with the diagonal elements.
Thus, this fixed point is IR attractive for $\xi<0$, $\eta<0$;

(1b) if $g'^*=\xi\frac{2(d-1)}{d-2+{\cal A}}$, $\Omega_{g'g'}^*=\xi$ and
$\Omega_{\alpha\alpha}^*=-\eta$, so that this fixed point is IR attractive
for $\xi>0$, $\eta<0$. For the remaining parameters $u$ and $f$ we have
the fixed-point values $u^*=({\cal A}-1)^2/(d-2+{\cal A})$ and
$f^*=\infty$ with $\Omega_{uu}^*=\Omega_{ff}^*=\xi$.

Another interesting case to be considered is $\alpha^*=\infty$.
From~\eqref{Dv} it follows that this case corresponds to the rapid-change
model with new charge $g''=g/\alpha^2$, which is supposed to be finite at
$\alpha\to\infty$. Besides that it is convenient to pass from the
variable $\alpha$ to variable $x=1/\alpha$, i.e., $x\to0$.
So, the new $\beta$-functions are
\begin{subequations}
\begin{equation}
\beta_{x}=x\eta;
\end{equation}
\begin{equation}
\beta_{u}=g''\left[\frac{(d-2+{\cal A})}{2(d-1)}u-
\frac{({\cal A}-1)^2}{2(d-1)}\right];
\end{equation}
\begin{equation}
\beta_f=g''\left[-f\frac{(d-2+{\cal A})}{2(d-1)}\right];
\end{equation}
\begin{equation}
\beta_{g''}=\frac{1}{\alpha^2}\beta_{g}-\frac{2g}{\alpha^3}\beta_{\alpha}
=g''\left[-\xi+\eta+\frac{(d-2+{\cal A})}{2(d-1)}g''\right].
\end{equation}
\end{subequations}

Thus, we find two more fixed points:

(2a) $g''^*=0$, with $\Omega_{g''g''}^*=-\xi+\eta$, $\Omega_{xx}^*=\eta$.
As  in the case (1a)
for two remaining parameters $u$ and $f$ we have $\beta_u=\beta_f\equiv0$,
$\Omega_{uu}^*=\Omega_{ff}^*\equiv0$, so both of them remain free
parameters.

As before the matrix $\Omega$ in the new variables
$\left\{g'',x,u,f\right\}$ is a matrix of the type~\eqref{Omega}, i.e., it
is triangle and its eigenvalues are simply given by
diagonal elements. Thus, this fixed point is IR attractive for
$\eta>0$, $\eta-\xi>0$;

(2b) if $g''^*=(\xi-\eta)\frac{2(d-1)}{d-2+{\cal A}}$,
$\Omega_{g''g''}^*=\xi-\eta$ and
$\Omega_{xx}^*=\eta$, so that this fixed point is IR attractive
for $\eta>0$, $\xi-\eta>0$. For the remaining parameters $u$ and $f$
we have the fixed-point values $u^*=({\cal A}-1)^2/(d-2+{\cal A})$
and $f^*=\infty$ with $\Omega_{uu}^*=\Omega_{ff}^*=\xi-\eta$.

For the special case $\eta = 0$ the function $\beta_\alpha$ and the
eigenvalue $\Omega_{\alpha\alpha}$ vanish identically,
so that the nontrivial fixed point
$\left[g/\alpha(\alpha+1)\right]^*=2\xi(d-1)/(d-2+{\cal A})$
is IR attractive for $\xi>0$.
Moreover, this fixed point is degenerate in the sense that we can not
determine the parameters $g^*$ and $\alpha^*$ separately.

\begin{figure}[h!]
\center
\includegraphics[width=0.47\textwidth,clip]{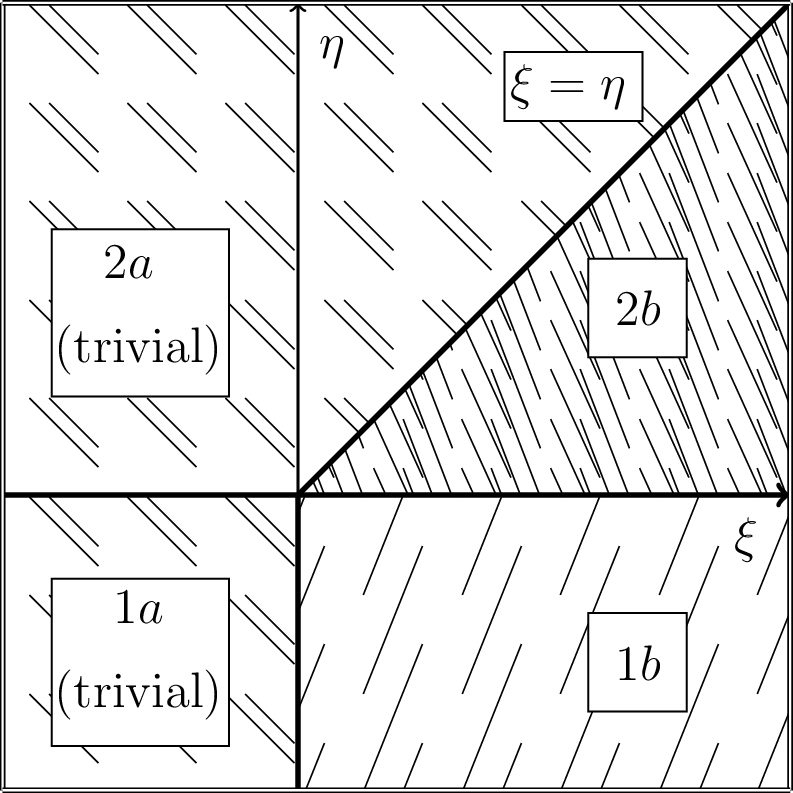}
\caption{Domains of IR stability of the fixed points in the model
(\protect\ref{Action}). The numbers in boxes correspond to the fixed
points (1{\it a})~--~(2{\it b}) in the text.}
%$P$ is the Kolmogorov point $\xi=2\eta=8/3$.}
\label{fig:Plain}
\end{figure}

Thus, we can conclude, that the domains of IR stability in this {\it vector}
model~\eqref{Action} coincide with the corresponding domains of IR stability
in {\it scalar} model, considered in~\cite{AntMal2011}.
The general pattern of the fixed points stability in the
$\xi$~---~$\eta$ plane is shown in Fig.~\ref{fig:Plain}.
The straight lines $\eta=0$; $\xi=0$, $\eta<0$; and $\xi=\eta$, $\eta>0$
corresponds to the boundaries of domains, which
has neither gaps nor overlaps between them.
Since the $\beta$-functions~\eqref{Beta} have no higher-order corrections,
this pattern is exact.

Note that the Kolmogorov values of the exponents
$\xi=8/3$, $\eta=4/3$ lie deep inside the domain of stability
of the nontrivial rapid-change point (2{\it b}); there is no borderline
going through this point.

This fact implies that the correlation functions of the model~(\ref{Action})
in the IR region ($\mu r\simeq\Lambda r \gg 1$, $Mr \sim 1$) exhibit scaling
behavior (as we will see below, up to logarithmic factors).

The corresponding critical dimensions $\Delta[F]\equiv\Delta_{F}$ for all
basic fields and parameters can be calculated exactly; see the next
subsection.

%%%%%%%%%%%%%%%%%%%%%%%%%%%%%%%%%%%%%%%%%%%%%%%%%%%%%%%%%%%%%%%%%%%%%%%%%%%%%%%%%%%%%%%%%%%%%%%%%%%%%%%%%%%%%%%%%%%%%%%%%%%%%%%%%%%%%%%%%%%%%%%%%%%%%%%%%%%%%%%%%%%%%%%%
\subsection{Critical dimensions}

In the leading order of the IR asymptotic behavior the Green functions
satisfy the RG equation~\eqref{RG-Eqtn} with the substitution
$g\to g^{*}$, $\alpha\to \alpha^{*}$, $f\to f^*$ and $u\to u^{*}$.
The operator ${\cal D}_{RG}$ is invariant with respect to the change
of variables $\left\{x,y\right\}\to\left\{x',y'\right\}$, i.e.,
$\beta_x\partial_x+\beta_y\partial_y=\beta_{x'}\partial_{x'}
+\beta_{y'}\partial_{y'}$. Taking into account the fact that $\gamma_u^*=0$,
this gives
\begin{equation}
\left[ {\cal D}_{\mu} - \gamma_{f}^{*}{\cal D}_{f}
+ \gamma_{G}^{*} \right] \,G^{R}(e,\mu,\dots) = 0.
\label{RGFP}
\end{equation}
Canonical scale invariance is expressed by the relations
\begin{equation}
\left[\sum _{\sigma}d_{\sigma}^k{\cal D}_{\sigma}-
d_{G}^k\right]G^{R}=0 ,\quad
\left[\sum _{\sigma}d_{\sigma}^{\omega }{\cal D}_{\sigma}-
d_{G}^{\omega }\right]G^{R}=0 ,
\label{Canonic-Scl-Inv}
\end{equation}
where $\sigma\equiv\{t,{\bf x},\mu,\nu,\alpha,m,M,u,f,{\cal A},g\}$ is the
set of all arguments of $G^{R}$ ($t,{\bf x}$ is the set of all times
and coordinates), and $d^{k}$ and $d^{\omega}$ are the
canonical dimensions of $G^{R}$ and $\sigma$. Substitution of the
needed dimensions from Table~\ref{table1}
and combination of the obtained result with~\eqref{RGFP} gives the
desired equation of critical IR scaling for the model:
\begin{equation}
\label{IR-Scaling}
\left[-{\cal D}_{\bf x}+ \Delta_{t} {\cal D}_{t} +
\Delta_{m} {\cal D}_{m} + \Delta_{M} {\cal D}_{M} +
\Delta_{f} {\cal D}_{f}-
\Delta_{G} \right]G^{R} = 0,
\end{equation}
where
$$ \Delta_{t}=-\Delta_{\omega}=-2, \quad
\Delta_{m} = \Delta_{M} =1, $$
\begin{equation}
\label{IR-Scaling-Coeff}
\Delta_{f} =\gamma_f^*,  \quad
\Delta_{u} =0
\end{equation}
and
\begin{equation}
\Delta[G]\equiv\Delta_{G} = d_{G}^{k}+ 2d_{G}^{\omega}+\gamma_{G}^{*}
\label{Critical-Dim}
\end{equation}
are the corresponding critical dimensions. Substituting the values of fixed
point of the regimes (1a)--(2b) we obtain:
\begin{eqnarray}
\Delta_f=0\quad\text{for (1a), (2a)}; & \\
\Delta_f=\xi\quad\text{for (1b)},&\quad\text{and}\quad
&\Delta_f=\xi-\eta\quad\text{for (2b)}.\nonumber
%\label{Critical-Dim}
\end{eqnarray}

In particular, for any correlation function
$G^{R}=\langle \Phi\dots\Phi\rangle$ of the fields $\Phi$
we have $\Delta_{G} = N_{\Phi} \Delta_{\Phi}$, with the summation
over all fields $\Phi$ entering into $ G^{R}$, namely,
\begin{equation}
\Delta_{G}= \sum_{\Phi} N_{\Phi}d_{\Phi} = N_{\theta'}d_{\theta'}+
N_{\theta}d_{\theta}+ N_{v}d_{v}.
\label{Ca}
\end{equation}
Since in the model~(\ref{Action}) the fields themselves are not renormalized
(i.e., $\gamma_{\Phi}=0$ for all $\Phi$,
see sec.~\ref{sec:TruePropagator}), using~(\ref{Critical-Dim}) we conclude,
that the critical dimensions of the fields $\Phi=\left\{\boldsymbol{v},
{\boldsymbol \theta},{\boldsymbol \theta'}\right\}$ are the same as their
canonical dimensions, presented in the Table~\ref{table1}. Namely,
\begin{equation}
\Delta_{\boldsymbol{v}}=1,\quad
\Delta_{\theta} = -1,\quad
\Delta_{\theta'} = d+1.
\label{33}
\end{equation}
It is the specific feature of the present model, which makes it similar
to the zero-correlation
time model~\cite{VectorN} and distinguishes it from
both the isotropic Kraichnan's vector model~\cite{AntGul2012-mod}
(in which $\gamma_\nu\neq0$) and anisotropic Kraichnan's scalar model
\cite{AntMal2011} (in which the Laplacian splitting parameter $f_0$ is
not dimensionless).

%%%%%%%%%%%%%%%%%%%%%%%%%%%%%%%%%%%%%%%%%%%%%%%%%%%%%%%%%%%%%%%%%%
\section{Renormalization and critical dimensions
of composite operators} \label{sec:Ops1}
%%%%%%%%%%%%%%%%%%%%%%%%%%%%%%%%%%%%%%%%%%%%%%%%%%%%%%%%%%%%%%%%%%

The analysis of the renormalization of composite operators is nearly
the same as in the rapid-change model~\cite{VectorN}, so we will discuss
it here very briefly.

\subsection{General scheme}
\label{sec:scheme}

The central role in the following will be played by composite fields
(``operators'') built solely of the basic fields $\theta$:
\begin{equation}
\label{F-N-p}
F_{Np}=(\theta_i\theta_i)^p\ (n_s\theta_s)^{2m},
\end{equation}
where $N=2(p+m)$ is the total number of fields $\theta$, entering
the operator.

As was pointed out in~\cite{VectorN}, the operator counterterms to a certain
$F_{Np}$ involve only operators of the form~\eqref{F-N-p} with the same value
of $N$. Besides that, all the corresponding diagrams diverge logarithmically
and one can calculate them with all external frequencies and momenta set
equal to zero.

Let us denote the closed set of operators, which can mix to each other in
renormalization, as $F\equiv\{F_{Np}\}$. The renormalization matrix
$\hat{Z}_{F}\equiv\{Z_{Np,Np'}\}$ for this set, given by the relation
\begin{equation}
%\label{gammaF-Def}
F_{Np}=\sum _{p'} Z_{Np,Np'} F_{Np'}^{R},
\end{equation}
is determined by the requirement that the 1-irreducible correlation function
\begin{widetext}
$$ \bigl\langle F_{Np}^{R} (x) \theta(x_{1})\dots\theta(x_{N})
\bigr\rangle_{\rm 1-ir}= $$
\begin{equation}
=\sum _{p'} Z_{Np,\,Np'}^{-1}\bigl\langle
F_{Np'}(x) \theta(x_{1})\dots\theta(x_{N})\bigr\rangle
_{\rm 1-ir} \equiv \sum _{p'}Z_{Np,\,Np'}^{-1}\Gamma_{Np'}
(x;x_{1},\dots, x_{N})
\end{equation}
\end{widetext}
be UV finite in renormalized theory, i.e., it has no poles in $\xi$ when
expressed in renormalized variables~(\ref{Renorm-Parameters}). This is
equivalent to the UV finiteness of the sum
$\sum_{p'}Z_{Np,\,Np'}^{-1}\Gamma_{Np'}(x;\theta)$, in which
$$\Gamma_{Np'} (x;\theta) = \frac{1}{N!}\, \int d x_{1}\dots \int d x_{N}\,
\Gamma_{Np'} (x;x_{1},\dots, x_{N})$$
\begin{equation}
\label{Gamma-Np}
\times\theta(x_{1})\dots\theta(x_{N})
\end{equation}
is a functional of the field $\theta(x)$.

The contribution of a specific diagram into the functional $\Gamma_{Np'}$
in~(\ref{Gamma-Np}) for any composite operator $F_{Np'}$ is represented in
the form
\begin{equation}
\label{Diag-General}
\Gamma_{Np'} = V_{\alpha\beta\dots} \, I^{ab\dots}_{\alpha\beta\dots} \,
\theta_{a} \theta_{b} \dots ,
\end{equation}
where $V_{\alpha\beta\dots}$ is the vertex factor,
$I^{ab\dots}_{\alpha\beta\dots}$ is the ``internal block''  of the diagram
with free vector indices, and the product $\theta_{a} \theta_{b} \dots$
corresponds to external ``tails.''

According to the general rules of the universal diagrammatic technique
(see, e.g.,~\cite{Vasiliev-Green}), for any composite operator $F(x)$
built of the fields $\theta$, the vertex
$V_{\alpha\beta\dots}$ in~(\ref{Diag-General})
with $k\ge0$ attached lines corresponds to the vertex factor
\begin{equation}
\label{Vertex-General}
V^{k}_{Np} (x;\, x_{1}, \dots, x_{k}) \equiv \delta^{k}
F_{Np}(x) / {\delta\theta(x_{1}) \dots\delta\theta(x_{k})}.
\end{equation}
The arguments $x_{1}\dots x_{k}$ of the quantity~(\ref{Vertex-General}) are
contracted with the arguments of the upper $\theta$ ends of the lines
$\langle\theta\theta'\rangle_{0}$ attached to the vertex.

%%%%%%%%%%%%%%%%%%%%%%%%%%%%%%%%%%%%%%%%%%%%%%%%%%%%%%%%%%%%%%%%%%%%%%%%%%%%%%%%%%%%%%%%%%%%%%%%%%%%%%%%%%%%%%%%%%%%%%%%%%%%%%%%%%%%%%%%%%%%%%%%%%%%%%%%%%%%%%%%%%%%%%%%
\subsection{Exact result for the diagrams}

\begin{figure}[h]
\center
\includegraphics[width=.22\textwidth,clip]{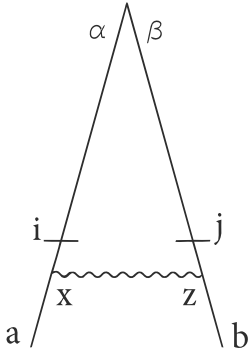}
\caption{The one-loop contribution to the generating
functional~(\ref{Gamma-Np}).}
\label{fig:One-Loop}
\end{figure}

Now let us turn to the calculation of the internal block
$I^{ab\dots}_{\alpha\beta\dots}$ of the diagrams.
The one-loop diagram is represented in Fig.~(\ref{fig:One-Loop}).
\noindent

Once all the external frequencies and momenta are set to zero,
the index structure of this diagram takes on the form
\begin{eqnarray}
%\begin{equation}
Y^{ab}_{\alpha\beta} &=& V_{xai}({\bf k})\ V_{zjb}({\bf -k})
P_{\alpha i}({\bf k})P_{\beta j}({\bf k})n_x n_z \nonumber \\
%\end{equation}
%\begin{equation}
&=&-{\cal A}^2n_xP_{x\alpha}({\bf k})n_zP_{z\beta}({\bf k})
\ k_ak_b,
%\end{equation}
\end{eqnarray}
where the letters $i,j,x$ and $z$ denote internal indices of the diagram
itself. Then we have to integrate $Y^{ab}_{\alpha\beta}$ over the frequency
and momentum with the factors like~\eqref{Dv}
and~(\ref{ThThAux-Propgr}), namely
\begin{eqnarray}
I^{ab}_{\alpha\beta} &=& \int\frac{d{\bf k}}{(2\pi)^d}\frac{1}
{-i\omega+\nu {\bf k}_\perp^2+\nu fk_\parallel^2}\frac{1}{i\omega+
\nu {\bf k}_\perp^2+\nu fk_\parallel^2} \nonumber \\
\label{Oper-1Loop-General}
&\times& 2\pi\delta(k_\parallel)\ D_0\
\frac{k_\perp^{5-d-(\xi+\eta)}}{\omega^2+\left[\alpha_0
\nu_0k_\perp^{2-\eta}\right]^2} \
Y^{ab}_{\alpha\beta}.
\end{eqnarray}

Since the expression~\eqref{Oper-1Loop-General} contains the factor
$\delta(k_\parallel)$, we can perform all the calculations with the original
propagators~\eqref{ThThAux-Propgr} and~\eqref{ThTh-Propgr}; see
the discussion in Sec.~\ref{sec:TruePropagator}.

Using the relation~(\ref{k-transv-Aver-Angles}) for averaging over the angles
and setting $\eta=0$ [see the discussion after~\eqref{Sigma-eta}],
we arrive at the following result:
\begin{eqnarray}
I^{ab}_{\alpha\beta} &=& \frac{{\cal A}^2f}{2\alpha(1+\alpha)} g
\int\frac{d{\bf k}_\perp}{(2\pi)^{d-1}}\
\frac{1}{k_\perp^{d-1+\xi}} \frac{k_a^\perp k_b^\perp}{k^2_\perp}\
n_\alpha n_\beta \nonumber \\
\label{1loop-diag-answer}
&=&\frac{{\cal A}^2 f}{2\alpha(\alpha+1)}\frac{1}{(d-1)} P_{ab}({\bf n})\
n_\alpha n_\beta\ g\times\frac{m^{-\xi}}{\xi}.
\end{eqnarray}

Contributions of all multiloop diagrams are equal to zero,
see~Sec.~\ref{sec:Multiloop}.
The multiloop diagrams of the ``sand clock'' type, represented
by products of simpler diagrams, contain only higher-order poles in
$\xi$ and, in the MS scheme, do not contribute to the anomalous dimensions.
Therefore the one-loop approximation~(\ref{1loop-diag-answer}) gives
the {\it exact} answer.

\subsection{Renormalization matrix and anomalous dimensions}
\label{sec:AmonDim}

Combining expressions~(\ref{Diag-General}),~(\ref{Vertex-General}) and
the exact answer~(\ref{1loop-diag-answer}),
for the functional $\Gamma_{Np}$ from~(\ref{Gamma-Np}) we obtain
\begin{widetext}
\begin{eqnarray}
\Gamma_{Np} &\propto&
\frac{\delta^2}{\delta\theta_\alpha\delta\theta_\beta}
\left[F_{Np}\right]\times n_\alpha n_\beta\times P_{ab}({\bf n})\times
\theta_a\theta_b =
\nonumber \\
&=& 2m(2m-1)\times F_{N\,p+1}\ +\ (2p+8pm-2m(2m-1))\times F_{N\,p}\ +
\nonumber \\
&+& (4p(p-1)-2p-8pm)\times F_{N\,p-1}\ -\ 4p(p-1)\times F_{N\,p-2},
\label{F-diag1}
\end{eqnarray}
\end{widetext}
up to an overall scalar factor.

Expression~(\ref{F-diag1}) shows that the operators $F_{Np}$ indeed
mix in renormalization: the UV finite renormalized operator $F^{R}$ has
the form $F^{R}=F+$ counterterms, where the contribution of the
counterterms is a linear combination of $F$ itself and other unrenormalized
operators with the same total number $N$ of the fields, which are said
to ``admix'' to $F$ in renormalization.

Let $F\equiv\{F_{p}\}$ be a closed set of operators (\ref{F-N-p}) with a
certain fixed value of $N$ (which we will omit below for  brevity) and
different values of $p$, which mix only to each other in renormalization.
The renormalization matrix $\hat{Z}_{F}\equiv\{Z_{p,p'}\}$
and the  matrix of anomalous dimensions
$\hat{\gamma}_{F}\equiv\{\gamma_{p,p'}\}$
for this set are given by
\begin{equation}
\label{gammaF-Def}
F_{p}=\sum _{p'} Z_{p,p'} F_{p'}^{R},
\quad
\hat{\gamma}_F=\hat{Z}_{F}^{-1}{\cal D}_{\mu }\hat{Z}_{F}.
\end{equation}
The scale invariance~(\ref{Canonic-Scl-Inv}) and the RG
equation~(\ref{RG-Eqtn}) for the operator $F_{p}$ give the corresponding
matrix of critical dimensions
$\Delta_{F}\equiv\{\Delta_{p,p'}\}$ in the form similar to the expression
(\ref{Critical-Dim}), where
$d_{F}^{k}$, $d_{F}^{\omega}$ and $d_{F}$ should be understood as the
diagonal matrices of canonical dimensions of the operators in
question (with the diagonal elements equal to sums of corresponding
dimensions of all fields and derivatives constituting $F$) and
$\hat{\gamma}^{*}=\hat{\gamma} (g^{*},\alpha^*,u^{*},f^*)$ is the matrix
(\ref{gammaF-Def}) at the fixed point.

In this notation and in the MS scheme the renormalization matrix $\hat{Z}$
has the form
\begin{equation}
\label{Z-F-Common}
\hat{Z} = \hat{I} + \hat{A},
\end{equation}
where $\hat{I}$ is the unity matrix and the elements of the matrix
$\hat{A}$ have the forms
\begin{equation}
\label{Z-diagram-matrix}
A_{pp'}=a_{pp'}\times \frac{g}{\xi}.
\end{equation}

Since the renormalization matrix $\hat{Z}$ has the form~(\ref{Z-F-Common}),
the matrix of anomalous dimensions $\hat{\gamma}$ has the form
\begin{equation}
\label{Gamma-Common-Aik}
\gamma_{pp'}=-a_{pp'}\ g
\end{equation}
with the coefficients $a_{pp'}$ from~(\ref{Z-diagram-matrix}). Combining
(\ref{F-diag1})~--~(\ref{Gamma-Common-Aik}) and taking into account the
scalar factor, not written in~(\ref{F-diag1}), but presented
in~(\ref{1loop-diag-answer}), together with the fact, that the symmetrical
coefficient for this one-loop diagram is $1/2$, one obtains the following
expression for the matrix of anomalous dimensions $\hat{\gamma}$:
\begin{eqnarray}
\label{gamma-F}
%\begin{align}
\gamma_{p,\,p'+1}&=&-\frac{{\cal A}^2 f}{4\alpha(\alpha+1)}\frac{1}{(d-1)}\
2m(2m-1)
\  g;
\nonumber \\
\gamma_{p,\,p'}&=&-\frac{{\cal A}^2 f}{4\alpha(\alpha+1)}\frac{1}{(d-1)}\
\left[2p+8pm-2m(2m-1)\right]
\  g; \nonumber \\
\gamma_{p,\,p'-1}&=&-\frac{{\cal A}^2 f}{4\alpha(\alpha+1)}\frac{1}{(d-1)}\
\left[4p(p-1)-2p-8pm\right]
\ g; \nonumber \\
\gamma_{p,\,p'-2}&=&-\frac{{\cal A}^2 f}{4\alpha(\alpha+1)}\frac{1}{(d-1)}\
\left[-4p(p-1)\right]
\ g.
%\end{align}
\end{eqnarray}

Now we have to substitute the value of the fixed point into the
expressions~\eqref{gamma-F}. For the critical regimes (1a) and (2a)
we immediately arrive at the trivial result $\gamma^*_{F}=0$. This means
that for such $\xi$ and $\eta$ the critical dimensions of the composite
operators coincide with their canonical dimensions, so that
there is no corrections to ordinary scaling.

For the regimes (1b) and (2b) we have $g'^*=\frac{2(d-1)}{d-2+{\cal A}}\xi$
and $g''^*=\frac{2(d-1)}{d-2+{\cal A}}(\xi-\eta)$, so that
\begin{eqnarray}
\label{gamma-F-fixed}
\gamma_{p,\,p'+1}^*&=&y\times
\ 2m(2m-1); \nonumber \\
\gamma_{p,\,p'}^*&=&y\times
\ \left[2p+8pm-2m(2m-1)\right]; \nonumber \\
\gamma_{p,\,p'-1}^*&=&y\times
\ \left[4p(p-1)-2p-8pm\right]; \nonumber \\
\gamma_{p,\,p'-2}^*&=&y\times
\ \left[-4p(p-1)\right],
\end{eqnarray}
where $y$ denotes the common factor, i.e.,
\begin{subequations}
\label{y}
\begin{equation}
y=-\frac{{\cal A}^2 f}{2(d-2+{\cal A})}\ \xi \quad
\text{for the critical regime (1b)};
\end{equation}
\begin{equation}
y=-\frac{{\cal A}^2 f}{2(d-2+{\cal A})}\ (\xi-\eta) \quad
\text{for the critical regime (2b)}.
\end{equation}
\end{subequations}

Therefore the matrix of critical dimensions for the set $F_{p}$ with fixed $N$
has the form
\begin{equation}
\label{Crit-Dim-F}
\Delta_{p,\,p'}=-2(p+m)\delta_{pp'}+\hat{\gamma}^*_{p,\,p'},
\end{equation}
where $-2(p+m)$ is the canonical dimension, $\delta_{pp'}$ is Kronecker's
$\delta$ symbol and $\hat{\gamma}^*_{p,p'}$ is the value of the matrix of
anomalous dimensions at the fixed point.

%%%%%%%%%%%%%%%%%%%%%%%%%%%%%%%%%%%%%%%%%%%%%%%%%%%%%%%%%%%%%%%%%%%%%%%%%%%%%%%%%%%%%%%%%%%%%%%%%%%%%
\subsection{Asymptotic behavior of the correlation function
$G=\left\langle F_1F_2 \right\rangle$}

Up to a scalar factor $y$, the values of the matrix elements of the matrix of
anomalous dimensions at the fixed point~\eqref{gamma-F-fixed} are the same as
in the zero-time correlation case~\cite{VectorN}.
This means that the matrix of critical dimensions~(\ref{Crit-Dim-F}) is not
diagonalizable, but can only be brought to the Jordan form, i.e.,
$\Delta_{F}= U_{F} \widetilde{\Delta}_{F} U_{F}^{-1}$, where the matrix
$\widetilde{\Delta}_{F}$ is
\begin{equation}
\label{Delta-F-Tilde}
\widetilde{\Delta}_{F}
=
\begin{pmatrix}
-2(p+m) & 1 & 0 & \dots & 0 \\
0 & -2(p+m) & 1 &  & \vdots \\
\vdots & 0 & \ddots & \ddots  &0 \\
\vdots &  &  & \ddots & 1 \\
0 & \dots &  & 0 & -2(p+m)
\end{pmatrix}.
\end{equation}

For the equal-time pair correlation function of two composite operators
$F_{Np}$ of the form~(\ref{F-N-p}) with arbitrary values of $N$ and $p$
\begin{eqnarray}
\label{G-Def}
G_{N_1p_1,\,N_2p_2}(r)=\left\langle F_{N_1p_1}(t,{\bf x}_{1}) \
F_{N_2p_2}(t,{\bf x}_{2}) \right\rangle,
\end{eqnarray}
where $r=|{\bf x}_{2}-{\bf x}_{1}|$,
$i=\left\{N_1p_1\right\}$, and $k=\left\{N_2p_2\right\}$,
this leads to the appearance of logarithmic dependence in the IR
asymptotic behavior (in the following we denote in $G_{ik}$ for brevity):
\begin{widetext}
\begin{equation}
\label{G-Asympt-General}
G_{ik}^R \propto (\mu r)^{N_1+N_2}
P_{(N_1+N_2)/2}\left[\ln\mu r\right]\ \Phi
\left(1,\,Mr,\,mr, \,\bar{f}\right) \ \ \ \ \forall i,k.
\end{equation}
\end{widetext}
Expression~\eqref{G-Asympt-General} is written up to a dimensional constant
factor; $P_L\left(\dots\right)$ is a polynomial of degree $L$ with the
argument $\ln \mu r$; $\bar{f}$ is the invariant charge and
$\bar{f}\to fr^\xi$ as $1/\mu r\to0$ for scaling regime (1b),
$\bar{f}\to fr^{\xi-\eta}$ as $1/\mu r\to0$ for scaling regime (2b).

Representations~(\ref{G-Asympt-General}) with yet unknown scaling functions
$\breve{\Phi}\left(Mr,\,mr,\,\bar{f}\right)\equiv\Phi\left(1,\,Mr,\,mr,
\,\bar{f}\right)$
describe the behavior of the correlation functions
for $\mu r\gg1$ and any fixed value of $Mr$. The inertial range
$\ell\ll r\ll L$  corresponds to the additional condition $Mr\ll 1$.
Here and below we do not distinguish the two
IR scales $M$ and $m$, first introduced in~\eqref{Cik} and~\eqref{VV};
the form of the functions
$\left.\breve{\Phi}\left(Mr,\,\bar{f}\right)\right|_{\bar{f}={\rm const}}$
as $Mr\to0$ is studied using the operator product expansion.

In general, the operators entering into the OPE are those which appear in the
corresponding Taylor expansions, and also all possible operators that admix
to them in renormalization~\cite{Zinn,Vasiliev-Green}. In our case the main
contribution to the sum is given by the operator
$F^R\propto (Mr)^{N_1+N_2} \times P_{(N_1+N_2)/2} \left[\ln Mr\right]$
which possesses maximal singularity.

Combining this fact with the RG representation~(\ref{G-Asympt-General}),
restoring canonical dimension $d_G=-N_1-N_2$ and retaining only the
leading term, we obtain the following asymptotic expression for the pair
correlation function $G$~(\ref{G-Def}) in the inertial range:
\begin{widetext}
\begin{equation}
\label{Answer}
G=\left\langle F_{N_1\,p_1}\ F_{N_2\,p_2} \right\rangle \propto
 \nu   ^{d^\omega_G} M^{-N_1-N_2} \left[\ln \mu r\right]^{(N_1+N_2)/2}
\left[\ln Mr\right]^{(N_1+N_2)/2}\ \widetilde{\Phi}
\left(\bar{f}\right),
\end{equation}
\end{widetext}
where $\widetilde{\Phi}\left(\bar{f}\right)$ is a certain scaling function,
restricted in the inertial range $\ell\ll r\ll L$. Owing to the nilpotency
of the matrix of critical dimensions,
the result obtained is independent of the scalar factor $y$~\eqref{y}, and
the only dependence on the exponents $\xi$ and $\eta$, that distinguishes
two nontrivial cases (1b) and (2b), is contained in the invariant charge
$\bar{f}$.

For the trivial regimes (1a) and (2a) there is no corrections to ordinary
scaling.

\section{Conclusion} \label{sec:Conc}

We applied the field theoretic renormalization group and the operator
product expansion to the analysis of the inertial-range asymptotic behavior
of a divergence-free vector field, passively advected by strongly
anisotropic turbulent flow.

Depending on the two exponents $\xi$ and $\eta$ that describe the
energy spectrum ${\cal E} \propto k_{\bot}^{1-\xi}$ and the dispersion
law $\omega \sim k_{\bot}^{2-\eta}$ of the velocity field,
the possible {\it nontrivial} types of the IR behavior appear to reduce
to only two limiting cases: the rapid-change type behavior, realized
for $\xi>\eta>0$, and the ``frozen'' (time-independent or ``quenched'')
behavior, realized for $\xi>0$, $\eta<0$.

To avoid possible confusion we stress that we studied the model with
arbitrary {\it finite} correlation time of the velocity field. The
behavior typical of the vanishing or infinite correlation time is
formed effectively in the IR range as the leading-order asymptotic
behavior  of the correlation functions.

In this respect, the situation is the same as in the model of the
anisotropic advection of
the {\it scalar} field, studied in~\cite{AntMal2011}. Thus, another
important conclusion of that work remains true~--
in contrast to the finite-correlated {\it isotropic} case,
where the Kolmogorov values $\xi/2=\eta=4/3$ lie exactly on the crossover
line between the
rapid-change and frozen regimes~\cite{FinTime,FinTimeEta,Chetak}, in the
present model they lie inside the domain of the rapid-change regime;
there is no crossover line going through this point. This result is in
agreement with the analysis of~\cite{Glimm-mod} and in disagreement
with the ~\cite{AM,AM1-mod} for the scalar case.

The inertial-range asymptotic expressions for various correlation functions
are summarized in expressions~(\ref{Answer}).
In contrast to the Kraichnan's rapid-change model, where the
correlation functions exhibit anomalous scaling behavior with infinite sets
of anomalous exponents, here the dependence on the integral turbulence
scale $L$ demonstrates a logarithmic character:
the anomalies manifest themselves as polynomials of logarithms of $(L/r)$,
where $r$ is the separation.

The key point is that the matrices of scaling dimensions of the relevant
families of composite fields (operators) appear nilpotent and cannot be
diagonalized~-- they can only be brought to Jordan form; hence the
logarithms.
This result is perturbatively exact in the sense that the contributions
of all multiloop diagrams appear equal to zero.

The possibility of logarithmic dependence of various correlation functions
on the integral scale $L$ and the separation $r$
should be taken into account in analysis of experimental data.
Since the difference between the nontrivial regimes (1b)
and (2b) stays only in the argument of the scaling function
$\widetilde{\Phi}$, it requires very accurate experiments to discern them.

It remains to admit that, although our model has a finite correlation time
and possess Galilean symmetry, it is still simplified in the sense that
the velocity ensemble is Gaussian. More realistic models should involve
the nonlinear NS equation, while the anisotropy should be introduced by
the large-scale stirring.
So far, the analysis based on the advecting NS velocity field was performed
only for the passive scalar \cite{NSpass} and vector \cite{Kotumay}
fields only in isotropic cases.

Thus, the analysis of the full-scale problem remains for the future;
this work is already in progress.

\section*{Acknowledgments}

The authors are indebted to L.~Ts.~Adzhemyan, Michal Hnatich, Juha Honkonen,
and S.~A.~Paston for discussions.

The work was supported by the Saint Petersburg State University within the
research grant 11.38.185.2014.
N.M.G. was also supported by the Dmitry Zimin's ``Dynasty'' foundation
and by the Saint Petersburg Committee of Science and High School.

%%%%%%%%%%%%%%%%%%%%%%%%%%%%%%%%%%%%%%%%%%%%%%%%%%%%%%%%%%%%%%%%%%%%%%%%%%%%%%%%%%%%%%%%%%%%%%%%%%%%%%%%%%%%%%%%%%%%%%%%%%%%%%%%%%%%%%%%%%%%%%%%%%%%%%%%%%%%%%%%%%%%%%%%
%%%%%%%%%%%%%%%%%%%%%%%%%%%%%%%%%%%%%%%%%%%%%%%%%%%%%%%%%%%%%%%%%%%%%%%%%%%%%%%%%%%%%%%%%%%%%%%%%%%%%%%%%%%%%%%%%%%%%%%%%%%%%%%%%%%%%%%%%%%%%%%%%%%%%%%%%%%%%%%%%%%%%%%%

\end{document}